\documentclass[amsmath,amssymb,nofootinbib,10pt,aps,prd,twocolumn,floatfix,showkeys,showpacs]{revtex4-2} 

\usepackage{graphicx}
\usepackage{subfig}
\usepackage{appendix}
\usepackage{xcolor} 
\usepackage{xfrac} 
\usepackage{here} 
\usepackage{courier} 

\usepackage{csquotes} 

\usepackage[colorlinks=true,linkcolor=blue,
            urlcolor=blue,citecolor=red,
            pdfusetitle]{hyperref}
\usepackage[nameinlink,capitalize]{cleveref}

\graphicspath{
    {./Figures}
}

\newcommand{\RN}[1]{
  \textup{\uppercase\expandafter{\romannumeral#1}}
}

\begin{document}
\title{The scalar angular Teukolsky equation and its solution for the Taub-NUT spacetime}
\author{Felix Willenborg}\email{felix.willenborg@zarm.uni-bremen.de} \author{Dennis Philipp}\email{dennis.philipp@zarm.uni-bremen.de} \author{Claus Lämmerzahl}\email{claus.laemmerzahl@zarm.uni-bremen.de}
\affiliation{
Zentrum für angewandte Raumfahrttechnologie und Mikrogravitation (ZARM), \\
Unversity of Bremen, 28359 Bremen, Germany
}
\affiliation{
Gauss-Olbers Space Technology Transfer Center, \\
c/o ZARM, Unversity of Bremen, 28359 Bremen, Germany
}

\begin{abstract}
\date\today 
The Taub-NUT spacetime offers many curious insights into the solutions of Einstein's electrovacuum equation. 
In the Bonnor interpretation, this spacetime possesses so-called Misner strings, which induce phenomena strikingly analogous to Dirac strings in the context of magnetic monopoles.
The study of scattering in the latter case leads to a quantization of the product of electric charge and magnetic moment, sometimes called the Dirac condition. 
To enable a thorough discussion of scattering on the Taub-NUT spacetime, linear perturbations are considered in the Newman-Penrose formalism and separated into angular and radial equations. 
The angular Teukolsky equation is discussed in detail, and eigenvalues are derived to subsequently solve the differential equation in terms of solutions to the confluent Heun equation. 
In the Bonnor interpretation of the Taub-NUT spacetime, there is no analog property to the Dirac condition. 
The choice of spacetime parameters remains unconstrained.
However, for a particular parameter choice, one can rederive the well-known \enquote{Misner} condition, in which a product of frequency and NUT charge is of integer value, as well as another product additionally including the Manko-Ruiz parameter.
The results of this work will allow us to solve analytically for wave-optical scattering in order to, e.g., examine the wave-optical image of Taub-NUT black holes. 
\end{abstract}



\maketitle
\tableofcontents

\section{Introduction}
General relativity predicts a range of observables by which a black hole can be characterized. 
Several experimental approaches have emerged, e.g. measurement of the motion of stars \cite{Abuter2020,Eckart1997} or gas clouds \cite{Gillessen2011} around Sgr A*, 
the Event Horizon Telescope observing gravitational lensing of the accretion disc \cite{EHTC2019,EHTC2022} or the LIGO detector measuring the gravitational radiation caused by merging black holes and their characteristic quasinormal modes in the ring-down phase \cite{LIGO2016}.
Such measurements can constrain the physical parameters of these compact objects, for example, the spin magnitude, the orientation of the spin axis, or the total mass.

Some black hole spacetimes, as electrovacuum solutions in general relativity, consider a larger number of physical properties than spin $a$ and mass $M$. 
One such parameter is the NUT charge $N$, related to the Taub-NUT spacetime \cite{Taub1951,Newman1963}.
This spacetime is rather unusual because it has many exotic properties not observed yet. 
For example, its bound timelike geodesics are limited to the surface of a cone or scattered lightlike geodesics are twisting the background for an observer in the context of gravitational lensing.
But even apart from these strange effects, the spacetime's interpretation is up for discussion as it contains conical singularities on its rotation axis - even outside of the horizon.
In a sense, the Taub-NUT spacetime shares many similarities with magnetic monopoles and their Dirac strings \cite{Fierz1944,Tamm1931,Wu1976,Nesterov2008}.
This leads to the coining of the NUT charge as the gravitomagnetic monopole. 
For example, the paths of charged particles around magnetic monopoles are similar to geodesics around Taub-NUT compact bodies: they move on cones. 
Moreover, formulating the spacetime in such a way that analogue gravitomagnetic and gravitoelectric fields arise allows a comparative discussion of both
\cite{LyndenBell1998,Nouri_Zonoz_1997,Bicak2000}. 
A striking property of magnetic monopoles is the quantization of the product of the electric charge and the magnetic monopole, called Dirac condition. 
This roots from the calculation based on the angular momentum operator of the charged particle around the magnetic monopole.

in the Taub-NUT spacetime, Misner \cite{Misner1963} addresses the issue of conical singularities by introducing specific coordinate patches of both hemispheres and considering a periodic time coordinate.
The period depends on the product of the NUT charge and the energy of a particle or rather the frequency of a wave.
Interestingly, the character of this \enquote{Misner} condition leads to a similar discretisation as in the Dirac condition.
However, the conical singularities vanish now completely in this approach, at the cost of introducing closed timelike curves through each event, possibly violating causality at a fundamental level.
Bonnor offers a different idea \cite{Bonnor1969}, consequently extended by Manko and Ruiz \cite{Manko2005}, in which they consider the conical singularity as a physical reality.
They can be seen as spinning rods with negative mass, inducing angular momentum into the spacetime but do not solve the presence of closed timelike curves. 
These appear in the proximity of the conical singularities, depending on the magnitude of the NUT charge.
Closed timelike curves cannot be eliminated in any interpretation of the Taub-NUT spacetime.
In the case of the Bonnor interpretation, even closed timelike and null geodesics can occur when exceeding a specific value of the Manko-Ruiz parameter \cite{Hennigar2019,Clement2015}.

Several examinations of the Taub-NUT spacetime and its phenomena are performed in the literature.
Gravitational lensing of geodesics in the Taub-NUT spacetime is considered in Refs. \cite{Jefremov2016,Kagramanova2010,Frost2022,Halla2023,Halla2020,Grenzebach2016}.
Besides these, the consideration of linear perturbations is another approach of examining the spacetime \cite{Bini2003,Rahman_2020,Lee2023,Kalamakis2021,Ciambelli2021}. 
This can either be done by linear perturbations in the metric or by reformulating the spacetime in the scope of the Newman-Penrose formalism \cite{Newman1962}. 
Teukolsky is one of the first to use this approach for the Kerr spacetime and introduced a rich tool for linear perturbations in general relativity, even allowing access to linear perturbations of higher spin-fields \cite{Teukolsky_1973,Press1973,Teukolsky1974}.
A master equation is derived in form of a linear second-order partial differential equation, which is separated into radial and angular ordinary differential equations. 

Differential equations of the type of the angular Teukolsky equation for the Kerr spacetime are related to spheroidal harmonics
\cite{Bouwkamp1947,Meixner1954,Blanch1946,Fackerell1977}, for Schwarzschild to the spherical harmonics, and for  generalizations of both to the spin-weighted spherical/spheroidal harmonics \cite{Goldberg1967,Newman1966}. Further generalizations in higher dimensions also exist \cite{Berti2006}. 
All these equations have in common that they are linear second-order ordinary differential equations with two regular singularities.
In case of the Kerr-de Sitter spacetime, the amount of regular singularities increases to four \cite{Willenborg2024,Suzuki_1998,Motohashi2021,Hatsuda2020}.
By this properties, the separated Teukolsky equations of the Plebański-Demiański are confluent or general Heun equations, respectively \cite{Batic_2007,Kamran1987}.
The solutions of such differential equations are known and discussed thoroughly \cite{Ronveaux1995}.
Using the analytical solutions of the separated Teukolsky equations enables analyzing scattering, quasinormal modes, transmission/absorption coefficients, metric reconstruction or even gray body factors \cite{Suzuki_1998,Suzuki_2000,Motohashi2021,Hatsuda2020,Kubota2024,Noda2022,Nambu2022,Hortacsu2020,Philipp2015,Schmidt2023,Willenborg2024,Berens2024}.
In a previous work \cite{Willenborg2024} we used the aforementioned methods to analytically treat wave-optical scattering in the Kerr-de Sitter spacetime, including the potential observations, cf. Refs.\ \cite{Nambu2016,Feldbrugge2020,Turyshev2020,Turyshev2021} for the ray-optical approach in gravitational lensing and  Refs.\ \cite{Frost2021,Frost2023,Frost2022,Frost2023a,Bohn2015,Perlick2022,Grenzebach2016}). 

The peculiar circumstances just mentioned justify a closer look at the angular Teukolsky equation of the Taub-NUT spacetime alone, before turning to the exact-analytical scattering of waves and the wave-optical observations in a future work. 
Here, we derive the solution of the angular Teukolsky equation and inspect it in the Misner as well as in the Bonnor interpretation.
Depending on the choice of interpretation, the parameters remain free or restricted to a condition.
This affects how multipole indices are used in the solutions and how completeness relations turn out, necessary for the calculation of scattering.
Besides the solution of the angular Teukolsky equation, the establishment of the completeness relation is also discussed, which is required in the calculation of the wave-optical scattering.

In the following, the spacetime and its properties in \cref{TN} are discussed.

\cref{TME} then deals shortly with the Taub-NUT metric in the Newman-Penrose formalism in order to derive the Teukolsky master equation. 
This can be solved by a separation Ansatz introducing arbitrary functions for the radial and the angular part and separating the temporal and azimuthal parts by exponentials due to symmetry properties of the manifold.

The separated equations of the Teukolsky master equations are \enquote{Heun equations in disguise} \cite{Suzuki_1998,Batic_2007}. 
In the case of the Taub-NUT spacetime, the radial and the angular part can be transformed to a confluent Heun equation, respectively.
We briefly discuss important properties of this differential equation and how to solve it in \cref{HeunC}.

With these introductory sections, we are able to derive the eigenvalue of separated Teukolsky equations by applying the methods to the angular equation. 
This is done in \cref{Eigenval}, and two approaches for the related spacetime interpretations are shown.

Finally, we give some solutions in \cref{Sol} as well as conclusions and an outlook in \cref{Conclusion}.

In the following we use geometric $G = c = 1$ units and the metric sign convention is $(+---)$. 

\section{Taub-NUT spacetime}
\label{TN}
The Taub-NUT spacetime is characterized by a mass parameter $M$, a NUT-charge $N$ and a Manko-Ruiz parameter $C$, where the metric can be written as\footnote{Note the sign for the $dt\,d\phi$ component. For example, compared to same metric sign convention in Ref. \cite{Bini2003,Jefremov2016} this differs. 
However, the metrics are isometric under the transformation $N \mapsto -N$. 
This difference simply redefines the direction in which the angular momentum is induced.
This can be seen in \cref{TN:Eq:metric} since $N$ appears squared everywhere except in the $dt\,d\phi$ component.
The same sign of the $dt\,d\phi$ component can be found in Ref. \cite{Kagramanova2010,Dowker1974} for same metric sign convention and for the inverted metric sign convention in \cite{Rahman_2020,Grenzebach2016,Halla2023,Griffiths2012}.
}
\begin{align}
    ds^2 =& \frac{\Delta_r}{\Sigma} dt^2 - \frac{\Sigma}{\Delta_r} dr^2 - d\theta^2 \Sigma - 2 \frac{\Delta_r \chi}{\Sigma} dt \,d\phi  \notag \\
    &+ \left(\frac{\Delta_r \chi^2}{\Sigma} - \sin^2\theta \Sigma\right) d\phi^2 \label{TN:Eq:metric} \, ,
\end{align}
written in terms of the functions of the Plebański–Demiański metric.
The exact forms can be found in \cref{App:PB}.
These reduce in the Taub-NUT case to
\begin{subequations}
    \begin{align}
        \Sigma &:= \Sigma(r) = r^2 + N^2 \, ,\\
        \Delta_r &:= \Delta_r(r) = r^2 - N^2 - 2 M r \, , \\
        \chi &:= \chi(\theta) = -2 N (C + \cos\theta) \, .
    \end{align}
\end{subequations}
It is convenient to know that $\Sigma = 1/(\rho \rho^*)$, with $\rho$ defined and used in \cref{App:NP:Eq:rho}. 

The parameter $\Sigma$ is non-zero everywhere for $N > 0$. 
Thus, with attention to \cref{TN:Eq:metric}, only the zeros of $\Delta_r$ in the coefficient of $dr^2$ determine the horizons of the Taub-NUT spacetime.
The event horizon is labeled $r_+$ and the Cauchy horizon $r_-$.
We get: 
\begin{subequations}
    \begin{align}
        r_+ &= M + \sqrt{M^2 + N^2} \\
        r_- &= M - \sqrt{M^2 + N^2}
    \end{align}
\end{subequations}
For the following distinct and real horizon radii are demanded.
The choice of parameters remains free at this stage, since there is no choice that leads to a naked singularity. 
Note, however, that in the case of a non-zero Kerr parameter $a$ or positive cosmological constant $\Lambda$ the choice of $N$ is restricted, see, e.g. Ref.\ \cite{Akcay_2011} in the case of Schwarzschild-de Sitter. 
Using the horizon radii, one can divide the radial regions into \cite{Griffiths2012}
\begin{itemize}
    \item domain of outer communication (NUT$_+$): $r_+ < r < \infty$: $\Delta_r > 0$ ,
    \item Taub region: $r_- < r < r_+$: $\Delta_r < 0$ ,
    \item NUT$_-$: $-\infty < r < r_-$: $\Delta_r > 0$ .
\end{itemize}

The first solution found was the inner Taub region, Taub's approach to a vacuum homogeneous cosmological model \cite{Taub1951}, which gave it its name.
Later on, Newman, Unti and Tamburino (NUT) derived a solution as a generalization of the exterior region, including the time-dependent inner Taub region in their representation \cite{Newman1963}.
Thus, the metric is usually called the Taub-NUT metric.

\begin{figure}[H]
        \centering
        \subfloat[$C < -1$]{\includegraphics[height=0.29\textheight]{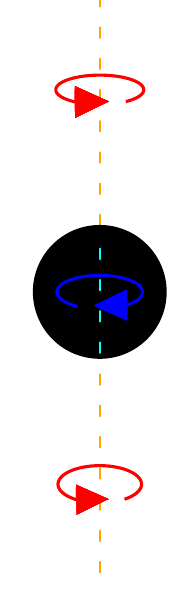}} \hfill
        \subfloat[$C = -1$]{\includegraphics[height=0.29\textheight]{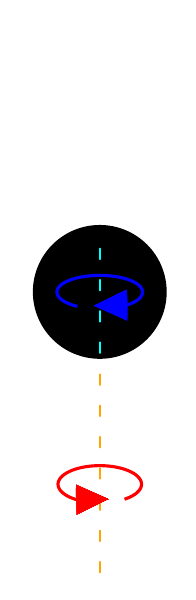}} \hfill
        \subfloat[$-1 < C < 0$]{\includegraphics[height=0.29\textheight]{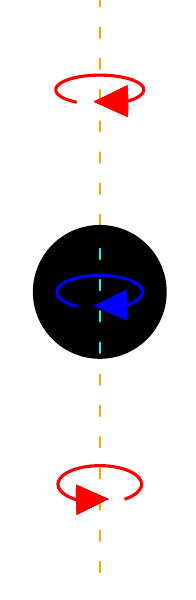}} \hfill
        \subfloat[$C = 0$]{\includegraphics[height=0.29\textheight]{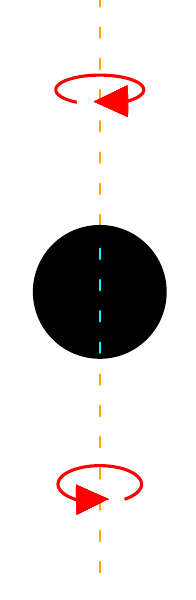}} \hfill
        \subfloat[$0 < C < 1$]{\includegraphics[height=0.29\textheight]{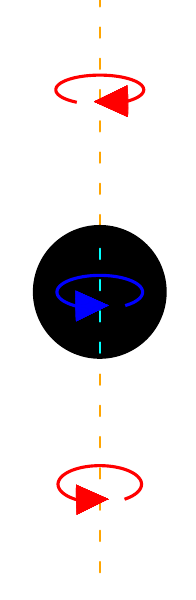}} \hfill
        \subfloat[$C = 1$]{\includegraphics[height=0.29\textheight]{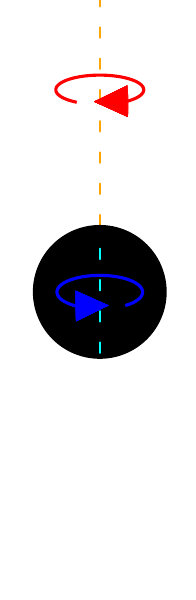}} \hfill
        \subfloat[$C > 1$]{\includegraphics[height=0.29\textheight]{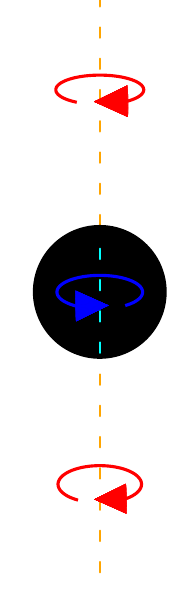}} \hfill
        \subfloat{\includegraphics[height=0.29\textheight]{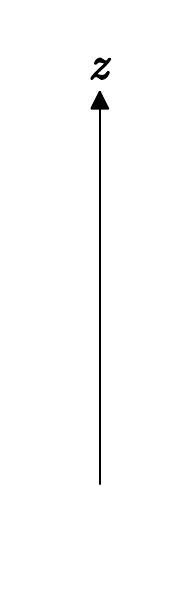}} \hfill
        \caption{Misner strings on the $z$-axis of the Taub-NUT black hole and their induced angular momentum depending on the choice of the Manko-Ruiz parameter $C$ for fixed Taub-NUT parameter $N$. Orange-dashed lines are the outer semi-infinite rods and cyan-dotted line is the inner rod. For $|C| = 1$ one of the semi-infinite strings, as well as their induced angular momentum disappears, while for $C = 0$ the inner string stays while its angular momentum vanishes (see Ref. \cite{Manko2005}).}
        \label{TN:Fig:StringAndMomentum}
    \end{figure}

The metric \cref{TN:Eq:metric} is curious in many senses. 
This led Misner even to call it a \enquote{counter-example to almost anything} \cite{Misner1967}. 
The major reason for this is the appearance of conical singularities on the rotational axis of the spacetime in the form of semi-infinite strings, even outside of the event horizon.
These singularities appear in the presence of a gravitomagnetic monopole ($N \neq 0$) and can be altered by the choice of $C$, the Manko-Ruiz parameter introduced in Ref. \cite{Manko2005}. 
There, exact representations can be found about the (Komar) masses and angular momenta of the semi-infinite strings.
For $|C| \neq 1$, the spacetime possesses two of these singularities as semi-infinite strings, and for $|C| = 1$ only one such string is present; see \cref{TN:Fig:StringAndMomentum} for reference. 
The choice of $|C| > 1$ causes even closed timelike geodesics, contrary to closed timelike curves otherwise \cite{Hennigar2019,Clement2015}.
Despite the presence of the strings, the choice of the parameter also influences the direction of the angular momentum and its magnitude.
Completely uninfluenced by this choice is the presence of the inner singularity string, located inside the event horizon and on the rotational axis as well. 
Its (Komar) mass is always nonzero, but its induced angular momentum vanishes for $C = 0$. 
The presence of the singularities makes the spacetime \cref{TN:Eq:metric} a non-asymptotically flat spacetime.

By a transformation $ x \mapsto - \bar{x}$ \cite{Jefremov2016}
\begin{subequations}
    \label{TN:Eq:EquatorialSymmetry}
    \begin{align}
        t &\mapsto \bar t \, , \\
        \phi &\mapsto -\bar \phi \, , \\
        \theta &\mapsto \pi - \bar \theta \, , \\
        r &\mapsto \bar r \, , 
    \end{align}
\end{subequations}
one can see that the resulting metric is not isometric to the original metric because $C$ is transformed to $\bar C = -C$.
This also shows that \cref{TN:Eq:metric} is not symmetric to the equatorial plane. 
But, the transformed Taub-NUT metric is globally isometric under this transformation to \cref{TN:Eq:metric} with $C \mapsto -C$.

Another transformation
\begin{subequations}
    \label{TN:Eq:LocalIsometry}
    \begin{align}
        t &\mapsto \bar t - 2 N (C - C') \, , \\
        \phi &\mapsto \bar \phi \, , \\
        \theta &\mapsto \bar \theta \, , \\
        r &\mapsto \bar r \, , 
    \end{align}
\end{subequations}
turns a Taub-NUT metric with $C$ into one with $C'$. 
Globally, these transformations are not valid until one assumes a periodic time coordinate $t$ with period $4 \pi |N (C - C')|$, see \cite{Jefremov2016}, because $\phi$ is a periodic coordinate.
Closed timelike curves (integral curves of $\partial_\phi$) in proximity to the singularity prevent a global isometry, thus limiting the transformation only to a local isometry in regions excluding these curves.
Misner \cite{Misner1963} considers the transformation \cref{TN:Eq:LocalIsometry} with $C = -1$ and $C - C' = 2$ and a global isometry, implying a periodic time coordinate with $T = 8 |N| \pi$.
The resulting spacetime contains a singularity on the other half of the axis, which he removed by patching the ill-behaved hemisphere with the original spacetimes regular hemisphere.
This results in a manifold with no singularity at all.
However, this is at the expense of every event now being on a closed timelike curve. 
In addition, a discretisation of $2 N \omega$ is introduced, which can be read from the time-dependent part of the separation (see \cref{Eigenval:AngMomOp}).

Bonnor reinterpreted these singularities as physical singularities, which are spinning semi-infinite rods with negative mass \cite{Bonnor1969,Manko2005}. 
These are sometimes also coined as Misner strings. 
The spinning rods induce angular momentum in the spacetime, as shown in refs. \cite{Nouri_Zonoz_1997,LyndenBell1998,Bicak2000,Griffiths2012}, where the spacetime was reformulated in an electromagnetic analogue with gravitomagnetic and gravitoelectric fields affecting the path of a massive particle. 
The resulting formulation brings the gravitomagnetic field in a form where it behaves as a monopole (radial alignment of fields lines) and where particles move on spatial cones.
This, together with the occurring strings, is very similar to the magnetic monopoles and the Dirac strings \cite{Fierz1944,Tamm1931,Wu1976,Nesterov2008}.
In the case of the Taub-NUT spacetime, the singularities have an effect on the spacetime, however, whereas the Dirac strings do not have an effect on the spacetime. 
This view on the Taub-NUT spacetime is called Bonnor interpretation.

The symmetries of the Taub-NUT spacetime lead to four linear independent Killing vector fields for any value of $M$, $N$ and $C$.
From \cref{TN:Eq:metric}, these are \cite{Halla2020,Halla2023}
\begin{subequations}
    \label{TN:Eq:KillingVectors}
    \begin{align}
        \xi_t &= \frac{\partial}{\partial t} \\
        \xi_x &= -\sin\theta \frac{\partial}{\partial \theta} - \frac{\cos\phi}{\sin\theta} \left(\cos\theta \frac{\partial}{\partial \phi} - 2 N (1 + C \cos\theta) \frac{\partial}{\partial t}\right) \\
        \xi_y &= \cos\theta \frac{\partial}{\partial \theta} - \frac{\sin\phi}{\sin\theta} \left(\cos\theta \frac{\partial}{\partial \phi} - 2 N (1 + C \cos\theta) \frac{\partial}{\partial t}\right) \\
        \xi_z &= \frac{\partial}{\partial \phi} - 2 N C \frac{\partial}{\partial t}
    \end{align}
\end{subequations}
where
\begin{subequations}
    \begin{align}
        [\xi_a, \xi_b] &= -\epsilon_{abc} \,\xi_c \\
        [\xi_t, \xi_a] &= 0
    \end{align}
\end{subequations}
and $a, b, c \in \{x, y, z\}$.

\onecolumngrid

\twocolumngrid
~\clearpage 

$\xi_a$ generate a three-dimensional group of isometries, which is isomorphic to the rotation group $SO(3, \mathbb{R})$, and $\xi_t$ generates a one-dimensional group of isometries that expresses stationarity. 
$N \neq 0$ leads to orbits of the rotations not coinciding with two-dimensional $S^2$ spheres anymore, but to submanifolds of $\mathbb{R} \times S^2$ with the signature $(+--)$\footnote{Note the sign invert here and in the Killing vector fields as previously mentioned due to our metric sign convention.}. 
A conclusion of this is that the Taub-NUT metric should not be called spherically symmetry, but as Ref. \cite{Halla2023,Halla2020} suggests \enquote{rotationally symmetric about any radial direction}. 
The Killing vector fields do not generally reflect the existence of a global symmetry group, but rather one of a local symmetry group due to the existence of the singularities, since they are not considered on them.
This is implied in the transformation \cref{TN:Eq:LocalIsometry}. 
Examining the Taub-NUT spacetime in the Misner interpretation, however, resolves this problem and recreates global symmetry \cite{Yohannes2021,Dowker1974}.
The latter property will become important in \cref{Eigenval:AngMomOp}.

\section{Teukolsky Equations}
\label{TME}
The Teukolsky Master Equation (TME) is a linear second-order partial differential equation for $t,r,\theta,\phi$, describing linear perturbations in the context of the Newman-Penrose formalism \cite{Teukolsky_1973,Teukolsky1974,Newman1962}. 
In \cref{App:NP} a short description is given of the most important steps leading to this equation, which for the Taub-NUT spacetime is
\begin{widetext}
    \begin{align}
    &\Delta_r^{-s} \frac{\partial}{\partial r}\left( \Delta_r^{s + 1} \frac{\partial \Psi}{\partial r} \right) 
    + \frac{1}{\sin\theta } \frac{\partial}{\partial \theta} \left(\Delta_\theta \sin\theta \frac{\partial \Psi}{\partial \theta} \right) 
    + \csc^2 \theta \frac{\partial^2 \Psi}{\partial \phi^2} + 2 i s \cot\theta \csc\theta \frac{\partial \Psi}{\partial \phi} + \left(\frac{\Sigma^2}{\Delta_r} - \csc^2 \theta ~\chi^2 \right) \frac{\partial^2 \Psi}{\partial t^2} \notag \\
    &+ s\left(\frac{4}{\rho^*} + 2 i \cot\theta \csc\theta \;\chi + \Sigma \frac{\Delta_r'}{\Delta_r}\right) \frac{\partial \Psi}{\partial t} + 2 \csc^2\theta \;\chi \frac{\partial^2 \Psi}{\partial t \partial \phi} - \left(s^2 \cot^2\theta - s\right) \Psi = 4 \pi \Sigma T \, , \label{TME:Eq:TME}
    \end{align}
\end{widetext}
where $\Psi := \Psi(t, r, \theta, \phi)$. 
The advantage of considering linear perturbations of all coefficients in the Newman-Penrose formalism instead of a intuitive linear-perturbation of the spacetime $\tilde{g}_{\mu\nu} = g{\mu\nu} + h_{\mu\nu}$, where $h_{\mu\nu}$ is the sought linear perturbation, is the access to different spin-fields $s$.
Check \cref{TME:Tab:NPGHPsol} for the implications of the choice of $s$. 
In this work, $s = 0$ is considered.
\cref{TME:Eq:TME} can be separated \cite{Kamran1987} and solved by the separation Ansatz 
\begin{align}
    \Psi(t, r, \theta, \phi) = R(r) S(\theta) e^{i m \phi} e^{-i \omega t} \, . \label{TMESep:Eq:SeparationAnsatz}
\end{align}
The underlying manifold introduces coordinates such that the $\phi$-coordinate is $2\pi$ periodic ($\phi \in [0, 2\pi] $). 
Therefore, $e^{i m \phi} = e^{i m (\phi + 2\pi)}$ and finally concluding that $m \in \mathbb{Z}$.

Inserting \cref{TMESep:Eq:SeparationAnsatz} into the TME gives two separated ordinary differential equations for $R(r)$ and $S(\theta)$.
The radial Teukolsky equation
\begin{align}
    \frac{d}{d r}\left(\Delta_r \frac{d}{d r} R(r)\right) + (V^\text{(rad)}(r) - \lambda) R(r) = 0
\end{align}
and the angular Teukolsky equation
\begin{align}
    \frac{1}{\sin\theta} \frac{d}{d\theta} \left(\sin\theta ~\Delta_\theta \frac{d}{d\theta} S(\theta)\right) + (V^\text{(ang)}_m(\theta) + \lambda) \,S(\theta) &= 0 \, . 
\label{TME:Sep:Eq:AngTheta}
\end{align}
For the coming discussion, only the angular Teukolsky equation will be of interest and the radial Teukolsky equation will be matter of a future publication.
Considering the separated differential equations together with their separation constant $\lambda$, latter is identified as an eigenvalue problem in the framework of the Sturm-Liouville theory.
The derivation of the eigenvalue is of importance in order solve our problem completely.
This will be achieved by transforming our angular Teukolsky equation into a different form.
A suitable differential equation is the confluent Heun equation, since the angular Teukolsky equation is a linear second-order differential equation with two regular singularities. 
In other words, the separated Teukolsky equations of the Plebański–Demiański spacetimes are \enquote{Heuns equations in disguise} (cf. \cite{Batic_2007}).
Before coming to the solution of \cref{TME:Sep:Eq:AngTheta} and the determination of its eigenvalue, we will thoroughly discuss the confluent Heun's equation and its later necessary properties.

\section{confluent Heun equation}
\label{HeunC}
The confluent Heun equation is a confluent form of the general Heun equation.
The canonical representation the confluent Heun equation is \cite{Ronveaux1995}
\begin{align}
    y''(z) + \left(\frac{\gamma}{z} + \frac{\delta}{z - 1} + \epsilon\right) y'(z) + \left(\frac{\alpha z - \sigma}{z (z - 1)}\right) y(z) = 0 \, . \label{HeunC:Eq:DEQ}
\end{align}
where $\sigma$ is the accessory parameter and $\alpha, \gamma, \beta, \epsilon$ are the exponent parameters, connected to the indicial exponents of their Frobenius expansion. 
At each regular singularity one can formulate two independent solutions, where at $z = 0$ the solutions have the indicial exponents $\{0, 1 - \gamma\}$ and at $z = 1$ they are $\{0, 1 - \delta\}$. 
In these solutions of respective regular singularities, $z = 0$ and $z = 1$ calculate to finite values. 
It is possible to derive solutions to this differential equation in two different ways: One uses a power series, giving the local confluent Heun function $CH\ell$ discussed in \cref{HeunC:CHl}. 
The property that distinguishes the $CH\ell$ from the other solutions to the differential equation is the amount of regular singularities in the convergence space.
Another solution uses favorably a function series Ansatz, by which it is possible to contain two regular singularities in a mutual convergence region.
These are called confluent Heun functions $CHf$ and are discussed in \cref{HeunC:CHf}.

\begin{widetext}
    \begin{subequations}
        \label{HeunC:CHl:Eq:CHl}
        \begin{align}
            y_{01}(z) &:= CH\ell(\sigma; \alpha, \gamma, \delta, \epsilon; z) \, , \label{HeunC:CHl:Eq:CHl:y01} \\
            y_{02}(z) &= z^{1 - \gamma} CH\ell(\sigma - (\delta - \epsilon) (1 - \gamma); \alpha + \epsilon (1 - \gamma), 2 - \gamma, \delta, \epsilon; z) \, , \label{HeunC:CHl:Eq:CHl:y02} \\
            y_{11}(z) &= CH\ell(\sigma - \alpha; -\alpha, \delta, \gamma, -\epsilon; 1 -z) \, , \label{HeunC:CHl:Eq:CHl:y11} \\
            y_{12}(z) &= (1 - z)^{1 - \delta} CH\ell(\sigma - (\gamma + \epsilon) (1 - \delta) - \alpha \epsilon; -(\alpha + (1 - \delta) \epsilon), 2 - \delta, \gamma, -\epsilon; 1 - z) \, . \label{HeunC:CHl:Eq:CHl:y12}
        \end{align}
    \end{subequations}
\end{widetext}

\subsection{local confluent Heun function \texorpdfstring{$CH\ell$}{CHl}}
\label{HeunC:CHl}
A power series Ansatz 
\begin{align}
    y(z) = \sum\limits^\infty_{n = 0} c_n x^n \label{HeunC:CHl:Eq:PowerSeries}
\end{align}
for a regular solution at $z = 0$, and is respective zeroth indicial exponent, will give us the local confluent Heun function. 
Inserting \cref{HeunC:CHl:Eq:PowerSeries} into \cref{HeunC:Eq:DEQ} results in the three-term recurrence relation
\begin{align}
    R_n c_{n+1} + S_n c_n + B_n c_{n-1} = 0 \, ,
\end{align}
where the coefficients are
\begin{subequations}
    \begin{align}
        R_n &= -(n + 1)(n + \gamma) \, ,\\
        S_n &= n (n - 1 + \gamma + \delta - \epsilon) - \sigma \, ,\\
        B_n &= \alpha + (n - 1) \epsilon \, ,
    \end{align}
\end{subequations}
with $c_0 = 1$ and $c_{-1} = 0$.
At each regular singularity $z \in \{0, 1\}$, two regular solutions can be formulated.
Thus, four in total.
With \cref{HeunC:CHl:Eq:PowerSeries} and $y(z) = y_{01}(z)$, all other regular solutions can be formulated by methods of automorphisms, see \cref{HeunC:CHl:Eq:CHl}.

The approximate behavior of solutions at $z = 0$ are
\begin{subequations}
    \label{HeunC:CHl:Eq:z0Approx}
    \begin{align}
        y_{01}(z) &= 1 + \mathcal{O}(z) \, , \\
        y_{02}(z) &= z^{1 - \gamma} \left(1 + \mathcal{O}(z)\right) \, ,
    \end{align}
\end{subequations}
and for $z = 1$
\begin{subequations}
    \label{HeunC:CHl:Eq:z1Approx}
    \begin{align}
        y_{11}(z) &= 1 + \mathcal{O}(1 - z) \, , \\
        y_{12}(z) &= (1 - z)^{1 - \gamma} \left(1 + \mathcal{O}(1 - z)\right) \, .
    \end{align}
\end{subequations}

The actual calculation of the $CH\ell$ can be done by many computer algebraic systems.
For our purpose, we rely on the calculation of the $CH\ell$ by the implementation in Mathematica \cite{WolframMathematicaHeunC2020}.

\subsection{confluent Heun function \texorpdfstring{$CHf$}{CHf}}
\label{HeunC:CHf}
Extending the region of convergence of the $CH\ell$ in order to contain both regular singularities leads to the construction of the confluent Heun function $CHf$.
For this, the Jacobi polynomials $P_n^{(\alpha, \beta)}$, being the solution of the differential equation
\begin{align}
    &(1 - x^2) y'' + \left(\beta - \alpha - (\alpha + \beta + 2) x\right) y' \notag \\ 
    &+ n (n + \alpha + \beta + 1) y = 0 \, ,
\end{align}
or rather
\begin{align}
    &z (z-1) y'' - (\beta + 1 - (2 + \alpha + \beta) z) y' \notag \\
    &- n (n + \alpha + \beta + 1) y = 0 \, , \label{HeunC:CHf:Eq:Jacobiz}
\end{align}
are chosen for our function series
\begin{align}
    y(z) = \sum\limits^\infty_{n = 0} g_n P_n^{(\gamma - 1, \delta - 1)}(z) \, ,\label{TMESol:Ang:Eq:JacobiFunSeries}
\end{align}
since the Jacobi equation is a special case of the confluent Heun equation. 
Again, $g_0 = 1$ and $g_{-1} = 0$. 
The convergence space expands to the whole complex plane of $z$ without $z = \infty$ instead of a limited region around the regular singularities.
By inserting \cref{TMESol:Ang:Eq:JacobiFunSeries} into \cref{HeunC:Eq:DEQ}, the derivation of the three-terms recurrence relation relies on the recurrence relations of the Jacobi polynomials, which are \cite{Ronveaux1995}
\begin{widetext}
    \begin{subequations}
        \label{HeunC:CHf:Eq:JacobiRecurrence}
        \begin{align}
            x P_{n}^{(\gamma - 1, \delta - 1)}(x) &= a_n P_{n + 1}^{(\gamma - 1, \delta - 1)}(x) + b_n P_{n}^{(\gamma - 1, \delta - 1)}(x) + c_n P_{n - 1}^{(\gamma - 1, \delta - 1)}(x) \, , \\
            (1 - x^2) \frac{d}{d x} P_{n}^{(\gamma - 1, \delta - 1)}(x) &= \tilde a_n P_{n + 1}^{(\gamma - 1, \delta - 1)}(x) + \tilde b_n P_{n}^{(\gamma - 1, \delta - 1)}(x) + \tilde c_n P_{n - 1}^{(\gamma - 1, \delta - 1)}(x) \, ,
        \end{align}
    \end{subequations}
\end{widetext}
with a coordinate transform $z = (x + 1)/2$, where the coefficients of the recurrence relations are listed in \cref{HeunC:CHf:Eq:JacobiRecurrenceCoeff} (cf. Ref. \cite[\href{https://dlmf.nist.gov/18.9}{§18.9}]{NIST:DLMF}).
This gives us the three-terms recurrence relation of the $CHf$
\begin{align}
    \label{HeunC:CHf:Eq:RecRel}
    \tilde R_n g_{n+1} + \tilde S_n g_n + \tilde B_n g_{n-1} = 0 \, ,
\end{align}
where $\tilde S_n = -(\tilde Q_n + \sigma)$ and finally
\begin{subequations}
    \begin{align}
        \label{HeunC:CHf:Eq:RecRelCoeff}
        \tilde R_n &= \epsilon \tilde c_{n+1} - \alpha c_{n+1} \, ,  \\
        \tilde S_n &= \sigma - n (n + \gamma + \delta + 1) - \alpha \left(b_n + \frac{1}{2}\right) + \epsilon \tilde b_n \, , \\
        \tilde B_n &= \epsilon \tilde a_{n-1} - \alpha a_{n-1} \, .
    \end{align}
\end{subequations}
Now, in order to turn $CH\ell$ finally into $CHf$, a property has to be imposed on the accessory parameter $\sigma$.
For this, the three-terms recurrence relation is transformed into the continued fractions
\begin{subequations}
    \begin{align}
        L_n &= -\frac{\tilde R_n}{\tilde S_n + B_n L_{n-1}} \, , \\
        M_n &= -\frac{\tilde B_n}{\tilde S_n + R_n M_{n+1}} \, .
    \end{align}
\end{subequations}
$M_n$ is an infinite continued fraction, whereas $L_n$ breaks at a certain point because of $g_{-1} = 0$.
Then, it holds that
\begin{align}
    M_n L_{n - 1} = 1 \, . \label{HeunC:CHf:Eq:ContinuedFrac}
\end{align}
By using $\tilde S_n = -\tilde Q - \sigma$ and rearranging for $\sigma$, \cref{HeunC:CHf:Eq:ContinuedFrac} yields
\begin{align}
    \sigma_k = -\frac{\tilde R_{n-1} \tilde B_n}{-\left(\tilde Q_{n-1} + \sigma_k \right) + \tilde B_{n-1} L_{n-2}} + \tilde R_n M_{n+1} - \tilde Q_n \, .\label{HeunC:CHf:Eq:EigenvalueEq}
\end{align}
Note that $\sigma_k$ gained an integer index $k$, denoting that now there is an infinite countable set of solutions giving confluent Heun functions $CHf$. 
Since $q_k$ appears also on the right hand side of the equation in the denominator of the continued fractions, it is solved in terms of a successive approximation (compare with the approach in \cite{Fackerell1977}). 

\begin{subequations}
    \label{HeunC:CHf:Eq:JacobiRecurrenceCoeff}
    \begin{align}
     a_n &= \frac{(n + 1) (n + \gamma + \delta - 1)}{(2n + \gamma + \delta - 1) (2n + \gamma + \delta)}, \\
     b_n &= \frac{(\gamma - \delta) (2 - \gamma - \delta)}{2 (2n + \gamma + \delta - 2) (2n + \gamma + \delta)}, \\
     c_n &= \frac{(n + \gamma - 1) (n + \delta - 1)}{(2n + \gamma + \delta - 2) (2n + \gamma + \delta - 1)}, \\
     \tilde a_n &= -\frac{n (n + 1) (n + \gamma + \delta - 1)}{(2n + \gamma + \delta -1) (2n + \gamma + \delta)}, \\
     \tilde b_n &= \frac{n (\gamma - \delta) (n + \gamma + \delta - 1) (\gamma - \delta)}{(2n + \gamma + \delta - 1) (2n + \gamma + \delta)}, \\
     \tilde c_n &= \frac{(n + \gamma - 1) (n + \delta - 1) (n + \gamma + \delta - 1)}{(2n + \gamma + \delta - 2) (2n + \gamma + \delta - 1)}.
    \end{align}
\end{subequations}

\subsubsection{Orthogonality and normalization}
The construction of a confluent Heun function $CHf$ implies the presence of orthogonality between solutions of different auxiliary parameters $\sigma_k$. 
This following calculation is closely related to the insights of Refs.  \cite{Ronveaux1995,Becker1997}.
There, it is written down for the Heun functions solutions $Hf$ of the general Heun equation. 
However, it finds equivalent application in the case of confluent Heun functions for the confluent Heun equation and the regular singularities $z \in \{0,1\}$.
Orthogonality in the case of confluent Heun functions is expressed by 
\begin{align}
    \left(q_k - q_n\right) \int_\mathcal{C} w(z) y_k(z) y_n(z) dz = \left[p(z) W_z\left[y_k, y_n\right]\right]_\mathcal{C} \, ,
    \label{HeunC:CHf:Normal:Eq:Ortho}
\end{align}
where 
\begin{subequations}
    \begin{align}
        w(z) 
        &= z^{\gamma - 1} (z - 1)^{\delta - 1} e^{\epsilon z} \, , \\
        p(z) 
        &= z^\gamma (z - 1)^\delta e^{\epsilon z} \, .
    \end{align}
\end{subequations}
$\mathcal{C}$ is a complex contour, generally constructed as a so-called \enquote{Pochhammer loop contour} around the regular singularities. 
However, Ref. \cite{Becker1997} has shown that a straight line connecting $z = 0$ and $z = 1$ also works out.
$y_k$ and $y_n$ are confluent Heun functions of same class with respective auxiliary parameters $\sigma_k$ and $\sigma_n$.
For different auxiliary parameter indices $k \neq n$, the right-hand side of \cref{HeunC:CHf:Normal:Eq:Ortho} becomes zero, but for $k = n$, this expression is non-zero:
\begin{align}
    \int_\mathcal{C} w(z) y_{ij}^2(q_k; z) dz = \zeta \, .
\end{align}
$\zeta$, as the result of this expression, will become important for normalizing the solutions derived from the confluent Heun functions in the calculation of scattering.
In Ref. \cite{Becker1997} a general expression of normalization constants for (confluent) Heun functions is given as
\begin{align}
    \zeta = -\Theta(z) \,p(z) \left.\frac{\partial W_z(z)}{\partial q}\right\rvert_{q=q_k} \, ,
    \label{HeunC:CHf:Normal:Eq:NormConst}
\end{align}
where 
\begin{subequations}
    \begin{align}
        W_z(z) &:= W_z\left[y_{01}(q, z), y_{11}(q, z)\right]\, , \\
        \Theta &:= \frac{y_{01}(q_k; z)}{y_{11}(q_k; z)} \, .
    \end{align}
\end{subequations}
Even though that $z$ is appearing as a variable in \cref{HeunC:CHf:Normal:Eq:NormConst}, $\zeta$ is independent of $z$ since $\Theta$ and $p(z) \left.\frac{\partial W_z}{\partial q}\right\rvert_{q=q_k}$ are independent of z in the mutual convergence space of $y_{01}(z)$ and $y_{11}(z)$ when they are confluent Heun functions.
We evaluate $\zeta$ in \cref{Sol} for $z = \sfrac{1}{2}$.

\section{Angular Teukolsky equation}
\label{AngTeuk}
By transforming $x = \cos\theta$ in \cref{TME:Sep:Eq:AngTheta}, one gets a more common known form of the differential equation:
\begin{align}
    \frac{d}{dx} \left(\Delta_x \frac{d}{dx} S(x)\right) + \left(\lambda - \frac{G_m^2(x)}{\Delta_x(x)}\right) \,S(x) &= 0 \, , \label{TMESol:Ang:Eq:DEQ}
\end{align}
where
\begin{subequations}
    \begin{align}
        G_m(x) &= m + 2 \omega N (C + x) \, , 
        \Delta_x = 1 - x^2 \, .
    \end{align}
\end{subequations}
For $N = 0$, \cref{TMESol:Ang:Eq:DEQ} is the differential equation of the spherical harmonics.
Note that now our points of interest, the coordinates of the poles, transform as $\theta \in \{0, \pi\} \mapsto x \in \{1, -1\}$.
The regular singularities of \cref{TMESol:Ang:Eq:DEQ} are the zeros of $\Delta_x(x) = 1 - x^2$:
\begin{align}
x_1 = -1 \, , \quad x_2 = 1 \, .
\end{align}
Besides the regular singularities, there is one irregular singularity at $x = \infty$.
By this properties already, one can identify \cref{TMESol:Ang:Eq:DEQ} as a Sturm-Liouville problem, in which $\lambda$ poses as an eigenvalue, and that its solutions are related to the solutions of the confluent Heun equation \cite{Ronveaux1995}.
A transformation of \cref{TMESol:Ang:Eq:DEQ} into the confluent Heun equation in two steps.
First, the independent variable $x$ is transformed by a Möbius transformation
\begin{align}
    \label{TME:Sep:Eq:IndependentVarTrafo}
    z = \frac{x + 1}{2} \, .
\end{align}
This transforms the coordinates of interest $x \in \{-1, 1, \infty\}$ to the regular singularities of the confluent Heun equation $z \in \{0, 1, \infty\}$.
The f-homotopic transformation on top of the dependent Variable $S(x)$ 
\begin{align}
    S(z) = z^{A_1} (z - 1)^{A_2} e^{A_3 z} y(z) \label{TME:Sep:Eq:DepTrans}
\end{align}
reduces the indicial exponents of the four regular solutions of \cref{TMESol:Ang:Eq:DEQ} such that they become $\{0, 2 A_1\}$ for solutions at $z = 0$ and $\{0, 2 A_2\}$ for solutions at $z = 1$.
\cref{TMESol:Ang:Eq:DEQ} now is 
\begin{widetext}
    \begin{align}
        \frac{d^2 y}{d z^2} + \left(\frac{2 A_1 + 1}{z} + \frac{2 A_2 + 1}{z - 1}\right)\frac{d y}{d z} - \frac{\lambda - (A_1 + A_2) (A_1 + A_2 + 1) + 4 N^2 \omega^2}{z (z -1)} y = 0.
    \label{TMESol:Ang:Eq:AngHeunEq}
    \end{align}
\end{widetext}
\cref{TMESol:Ang:Eq:AngHeunEq} is now the confluent Heun equation in a different form.
By comparing the coefficients of \cref{TMESol:Ang:Eq:AngHeunEq,HeunC:Eq:DEQ}, one gets
\begin{subequations}
    \begin{align}
        \gamma &= 2 A_1 + 1 \\
        \delta &= 2 A_2 + 1 \\
        \epsilon &= 0 \\
        \alpha &= 0 \\
        \sigma &= \lambda + (A_1 + A_2) (A_1 + A_2 + 1) - 4 N^2 \omega^2
    \end{align}
    \label{TMESol:Ang:Eq:Coefficients}
\end{subequations}
The exponents $A_1$, $A_2$ will have to take a particular form in order for the transformation to the confluent Heun equation to work out, which is
\begin{align}
    A_i^2 &= A^2(x_i) \, ,
\end{align}
with
\begin{align}
    A(x) &= \frac{G_m(x)}{\Delta'_x (x)}
\end{align}
and $i \in \{1, 2\}$. 
Thus, explicitly written out the exponents take the form
\begin{subequations}
    \label{TMESol:Ang:Eq:Exponent}
    \begin{align}
        A_1 &= \pm \left|\frac{m + 2 N \omega (C - 1)}{2}\right| \, , \\
        A_2 &= \pm \left|\frac{m + 2 N \omega (C + 1)}{2}\right| \, .
    \end{align}
\end{subequations}
The remaining exponent turns out to be $A_3 = 0$. 
What is left is a sign ambiguity, which has to be resolved by a boundary condition. 
For our solutions we demand regularity on both poles. 
This is motivated by preserving the limit $N \rightarrow 0$ and the known solutions there, and by creating a complete orthonormal set of functions. 
The latter is required, for example, in the calculation of scattering (cf. e.g. \cite{Willenborg2024}). 
A complete set of orthonormal functions implies a finite normalization constant as in \cref{HeunC:CHf:Normal:Eq:NormConst}. 
For a diverging solution, this would be infinite and not applicable anymore. 
If we did not have a complete set of orthonormal functions, the expansion of the radial Teukolsky equation in terms of the solution of the angular Teukolsky equation would no longer be possible. 

Looking at all solutions and their approximate behavior at their respective regular singularities, we can use \cref{HeunC:CHl:Eq:z0Approx,HeunC:CHl:Eq:z1Approx} to find
\begin{subequations}
    \begin{align}
        S_{01}(x) &\propto (1 + x)^{A_1} \left(1 + \mathcal{O}(1 + x)\right) \, , \\
        S_{02}(x) &\propto (1 + x)^{-A_1} \left(1 + \mathcal{O}(1 + x)\right) \, , \\
        S_{11}(x) &\propto (1 - x)^{A_2} \left(1 + \mathcal{O}(1 - x)\right) \, , \\
        S_{12}(x) &\propto (1 - x)^{-A_2} \left(1 + \mathcal{O}(1 - x)\right) \, .
    \end{align}
\end{subequations}
Here we have four cases we can choose from:
\begin{itemize}
    \item $A_1$, $A_2$ positive signs: This case requires that only $S_{01}(x)$ and $S_{11}(x)$ are solutions that satisfy our regularity condition.
    \item $A_1$, $A_2$ negative signs: This case requires that only $S_{02}(x)$ and $S_{12}(x)$ are solutions that satisfy our regularity condition. 
    \item $A_1$, $A_2$ having positive and negative signs respectively: Depending on the parameters $m$, $N$, $\omega$, and $C$ we have to switch from one solution $S_{01}(x)$ to $S_{02}(x)$, which also holds for the solutions at $z = 1$. See Ref. \cite{Motohashi2021,Hatsuda2020,Willenborg2024} for such an application
    \item $A_1$, $A_2$ having negative and positive signs respectively: Same description as before but with inverted signs
\end{itemize}
The functions chosen to be our solutions of the angular Teukolsky equation and regular at the regular singularities are analog (and also limiting against for the case $N \rightarrow 0$) to the Legendre functions of the first kind $P_n(x)$.
We want our derivation to be closely related to the derivation of the already known spherical harmonics and spheroidal harmonics \cite{Fackerell1977}, as well as their spin-weighted generalisations \cite{Berti2006,Goldberg1967}. 
Thus, we choose positive signs resulting in $S_{01}(x)$ and $S_{11}(x)$ being our relevant functions.
Therefore, the exponents take the final form
\begin{subequations}
    \label{TMESol:Ang:Eq:FinalExponent}
    \begin{align}
        A_1 &= \left|\frac{m + 2 N \omega (C - 1)}{2}\right| \, , \\
        A_2 &= \left|\frac{m + 2 N \omega (C + 1)}{2}\right| \, ,
    \end{align}
\end{subequations}
giving us the final solution of the angular Teukolsky equation:
\begin{align}
    S(x) = c_1 S_{01}(x) + c_2 S_{11}(x)
\end{align}

All coefficients for this solutions are completely determined except the accessory parameter $\sigma$, containing the yet undetermined eigenvalue $\lambda$. 
The determination follows the Sturm-Liouville theory, since the separation of the Teukolsky master equation results in a Sturm-Liouville problem of the eigenvalue. 
We want to discuss the eigenvalues in closer detail.

\section{Eigenvalue problem}
\label{Eigenval}
The determination of the eigenvalue for the solutions shall now be of main focus, since important properties can be derived here.
One can determine the eigenvalue by two approaches: By using the three-terms recurrence relation \cref{HeunC:CHf:Eq:EigenvalueEq} or by using angular momentum operators, e.g. in the case of spherical harmonics, spheroidal harmonics and their spin-weighted generalisations. 
It will turn out in order to completely solve this problem, a hybrid approach of both approaches will be of importance, depending on the interpretation used.
This is why we will discuss both now, starting with the three-terms recurrence relation approach.

\subsection{Eigenvalue by recurrence relation}
\label{Eigenval:Recurrence}
As already mentioned, we will use \cref{HeunC:CHf:Eq:EigenvalueEq} together with the determined coefficients \cref{TMESol:Ang:Eq:Coefficients} for the confluent Heun functions. 
The motivation for this is that for each solution regularity at both poles simultaneously is desired.
This will give us the form of the eigenvalue by rearranging for $\lambda$:
\begin{align}
    \lambda = \, &n (n + \gamma + \delta - 1) + \frac{\gamma + \delta - 2}{2} \left(\frac{\gamma + \delta - 2}{2} + 1\right) \notag \\
    & + 4 N^2 \omega^2 \label{Eigenval:Recurrence:Eq:BonnorEigenval}
\end{align}
Inserting \cref{Eigenval:Recurrence:Eq:BonnorEigenval} into \cref{TMESol:Ang:Eq:AngHeunEq} turns the differential equation into a form such that it coincides with the Jacobi equation \cref{HeunC:CHf:Eq:Jacobiz}. 
This is due to $\epsilon = \alpha = 0$, which expresses that the solution by confluent Heun equation coincide with the Jacobi equation and the function series \cref{TMESol:Ang:Eq:JacobiFunSeries} decays into a single Jacobi function $P_n^{(\alpha, \beta)}(z)$. 
It should be emphasized that in the case of a non-zero Kerr parameter $a$ that $\epsilon \neq 0$, $\alpha \neq 0$. 
The derivation of the solution of the angular Teukolsky equation will therefore come in handy in a future work including a non-zero Kerr parameter.

Redefining $n$, thus that it coincides with the approach for spherical, spheroidal, spin-weighted spherical or spin-weighted spheroidal functions \cite{Fackerell1977,Berti2006,Goldberg1967}
\begin{align}
    n := l - \frac{\gamma + \delta - 2}{2} \, ,\label{TMESol:Ang:Eq:Redefinition}
\end{align}
will lead to a form from which the limiting behavior $N \rightarrow 0$ is easily read off:
\begin{align}
    \boxed{
    \lambda = l(l + 1) + 4 N^2 \omega^2
    } \label{Eigenval:Recurrence:Eq:Result}\, .
\end{align}
This closely resembles the known eigenvalue for (spin-weighted) spherical harmonics, except for an additional $4 N^2 \omega^2$. 
Finally, all unknown elements are solved for the angular Teukolsky equation.
When inserting \cref{Eigenval:Recurrence:Eq:Result} into \cref{TMESol:Ang:Eq:DEQ} the form of \cref{TMESol:Ang:Eq:FinalForm} is yielded, relevant in \cref{Eigenval:AngMomOp}.
\begin{widetext}
    \begin{align}
        \frac{d}{d x}\left((1 - x^2) \frac{d}{d x} S(x)\right) - \left(\frac{(m + 2 N \omega (C - 1))^2}{2 (1 + x)} + \frac{(m + 2 N \omega (C + 1))^2}{2 (1 -x)} - l(l+1)\right) S(x) &= 0 \label{TMESol:Ang:Eq:FinalForm}
    \end{align}
\end{widetext}

We can simplify this expression furthermore using the fact that our solutions are confluent Heun functions. 
Since a solution at $z = 0$ is now regular at $z \in \{0, 1\}$, a solution at $z = 1$ is also regular at $z \in \{0, 1\}$, both solutions are linearly dependent to each other.
Thus, one solution can be expressed by the other one with an additional proportional constant
\begin{align}
    \label{TMESol:Ang:Eq:PropConst}
    \Theta_{ij} = \frac{S_{0i}(0)}{S_{1j}(0)} \, .
\end{align}
\cref{TMESol:Ang:Eq:PropConst} is independent on $x$ and evaluated at an arbitrary point in the mutual convergence space. 
Here, $x = 0$ is chosen. 
The final solution simplifies through the linear dependency relation to
\begin{align}
    S(x) = \left(c_1 + \frac{c_2}{\Theta_{11}}\right) S_{01}(x) \, .
\end{align}
We can rescale $S(x)$ such that the constant vanishes, and divide our solution by the normalization constant $\zeta$ in \cref{HeunC:CHf:Normal:Eq:NormConst} for confluent Heun functions in order to turn our solutions into a set of complete orthonormal functions
\begin{align}
    \tilde S(x) = \frac{1}{\left(c_1 + \frac{c_2}{\Theta_{11}}\right) \sqrt{\zeta}} S(x) \label{TMESol:Ang:Eq:FinalSolution}
\end{align}
where the choice of $c_1$ and $c_2$ becomes obsolete, since these are global scaling factors.
\cref{TMESol:Ang:Eq:FinalSolution} is now the final solution of the angular Teukolsky equation for the Taub-NUT spacetime. 

Establishing a complete orthonormal set of functions comes with the benefit of gaining another property: the completeness relation of our solutions. 
In which turns out in this case as
\begin{align}
    \sum\limits^\infty_{n = 0} \sum\limits^\infty_{m = -\infty} \tilde S_{nm}(x) \tilde S_{nm}(x') = \delta(x - x')\, . \label{TMESol:Ang:Eq:CompletenessRelationBonnor}
\end{align}
It will become important when calculating wave-optical scattering by a Greens function (cf. Refs. \cite{Willenborg2024,Motohashi2021}). 

Let us return to the redefinition of the running index $n$, since it is unusual in the topic of angular harmonics to use $n$ instead of an angular momentum number $l$. 
The consequence of this choice reveals itself by looking closer at the sum of $\gamma$ and $\delta$:
\begin{subequations}
    \begin{align}
        \gamma + \delta - 2 &= 2 \max(|2 N \omega|,|m + 2 N \omega C|)
    \end{align}
\end{subequations}
Inserting this into \cref{TMESol:Ang:Eq:Redefinition} and rearranging for the index $l$ yields
\begin{align}
    l_n = n + \max(|2 N \omega|,|m + 2 N \omega C|) \, .\label{TMESol:Ang:Eq:lRule}
\end{align}
Such a relation can be derived for spherical and spheroidal harmonics, as well as for their spin-weighted generalisations from the recurrence relations. 
From \cref{TMESol:Ang:Eq:lRule}, and the fact that $n \in \mathbb{N}_0$, we see that $l \geqslant \max(|2 N \omega|,|m + 2 N \omega C|)$.
We have $l \geqslant \max(|m|, |s|)$ for the spin-weighted spherical and spheroidal harmonics \cite{Goldberg1967} and $l \geqslant |m|$ for the non-spin-weighted special cases \cite{Fackerell1977,Berti2006}.

In case of the spherical harmonics ($s = N = 0$) we know that $n = l - |m|$. 
The possible values can be represented as follows for different $m$ and $n$:
\begin{table}[H]
    \centering
    \begin{tabular}{ccccccccccccc}
         {} & {} & {} & {} & {} & {} & {\color{red} $(0, 0)$} & {} & {} & {} & {} & {} & {}  \\
         {} & {} & {} & {} & {} & {\color{red} $(1, -1)$} & {\color{green} $(1, 0)$} & {\color{red} $(1, 1)$} & {} & {} & {} & {} & {} \\
         {} & {} & {} & {} & {\color{red} $(2, -2)$} & {\color{green} $(2, -1)$} & {\color{blue} $(2, 0)$} & {\color{green} $(2, 1)$} & {\color{red} $(2, 2)$} & {} & {} & {} & {}
    \end{tabular}
\end{table}
where the elements are tuples $(l, m)$.
The colors represent the cases ${\color{red} n = 0}$, ${\color{green} n = 1}$, ${\color{blue} n = 2}$, $\ldots$. 
For constant $n$, the states form a triangular line in this representation. 
Increasing $n$ by one shifts the triangle one row downwards.
Generalizing to the spin-weighted spherical/spheroidal harmonics, where the spin-weight $s \in \mathbb{Z} / 2$, e.g $|s| = \sfrac{1}{2}$, the possible values for $l = n - \max(|m|, |s|)$ are
\begin{table}[H]
    \centering
    \begin{tabular}{ccccccc}
         {} & {} & {\color{red} $(\sfrac{1}{2}, -\sfrac{1}{2})$} & {\color{red} $(\sfrac{1}{2}, \sfrac{1}{2})$} & {} & {} \\
         {} & {\color{red} $(\sfrac{3}{2}, -\sfrac{3}{2})$} & {\color{green} $(\sfrac{3}{2}, -\sfrac{1}{2})$} & {\color{green} $(\sfrac{3}{2}, \sfrac{1}{2})$} & {\color{red} $(\sfrac{3}{2}, \sfrac{3}{2})$} & {} \\
         {\color{red} $(\sfrac{5}{2}, -\sfrac{5}{2})$} & {\color{green} $(\sfrac{5}{2}, -\sfrac{3}{2})$} & {\color{blue} $(\sfrac{5}{2}, -\sfrac{1}{2})$} & {\color{blue} $(\sfrac{5}{2}, \sfrac{1}{2})$} & {\color{green} $(\sfrac{5}{2}, \sfrac{3}{2})$} & {\color{red} $(\sfrac{5}{2}, \sfrac{5}{2})$}
    \end{tabular}
\end{table}
The difference here is $l, m \in \mathbb{Z} / 2 \backslash \mathbb{Z}$ and two cases, where $|s| \geqslant |m|$ and $|m| > |s|$ for fixed $|s|$. 
In the first case we can have two tuples, contrary to the non-spin-weighted case where there is only one.
The second case has again infinite countable tuples.
For $|s| = 1$
\begin{table}[H]
    \centering
    \begin{tabular}{ccccccc}
         {} & {} & {\color{red} $(1, -1)$} & {\color{red} $(1, 0)$} & {\color{red} $(1, 1)$} & {} & {} \\
         {} & {\color{red} $(2, -2)$} & {\color{green} $(2, -1)$} & {\color{green} $(2, 0)$} & {\color{green} $(2, 1)$} & {\color{red} $(2, 2)$} & {} \\
         {\color{red} $(3 -3)$} & {\color{green} $(3, -2)$} & {\color{blue} $(3, -1)$} & {\color{blue} $3, 0)$} & {\color{blue} $(3, 1)$} & {\color{green} $(3, 2)$} & {\color{red} $(3, 3)$}
    \end{tabular}
\end{table}
this goes even further, where three tuples fulfill $|s| = |m|$ for fixed $|s|$.
In order to examine this behavior for the solution of the angular Teukolsky equation, and assuming that we can choose our parameters freely, where $\omega, N > 0$ and $C \in \mathbb{R}$, we choose as an example $C = 0$, $N = 0.32M$ and $\omega = 2.275 / M$ ($2 N \omega = 1.456$).
An equivalent set of indices would take the form of
\begin{table}[H]
    \centering
    \begin{tabular}{ccccccc}
        {} & {} & {\color{red} $(l_0, -1)$} & {\color{red} $(l_0, 0)$} & {\color{red} $(l_0, 1)$} & {} & {} \\
        {} & {\color{red} $(2, -2)$} & {\color{green} $(l_1, -1)$} & {\color{green} $(l_1, 0)$} & {\color{green} $(l_1, 1)$} & {\color{red} $(2, 2)$} & {} \\
        {\color{red} $(3, -3)$} & {\color{green} $(3, -2)$} & {\color{blue} $(l_2, -1)$} & {\color{blue} $(l_2, 0)$} & {\color{blue} $(l_2, 1)$} & {\color{green} $(3, 2)$} & {\color{red} $(3, -3)$} 
    \end{tabular}
\end{table}
Again, we have two different cases $|2 N \omega| \geqslant |m + 2 N \omega C|$ and $|m + 2 N \omega C| > |2 N \omega|$. 
Up until now we covered the strict and mathematical derivation of the eigenvalue of the differential equation \cref{TMESol:Ang:Eq:DEQ} in terms of the recurrence relation. 
Physically, the derivation is applicable to both interpretations of the spacetime. 
For the time being, it seems that we can freely chose any physical parameter $M, N, C$. 
By these we see that $l$ is derived by \cref{TMESol:Ang:Eq:lRule} and from \cref{TMESep:Eq:SeparationAnsatz} we know that $m \in \mathbb{Z}$. 
Thus, $l \in \mathbb{R}$ generally instead of $l \in \mathbb{Z} / 2$. 

What is left out is that the (spin-weighted) spherical/spheroidal harmonics profit from the approach via angular momentum operators. 
The derivation of the eigenvalue and the solution of the angular Teukolsky equation can also be done in these. 
It is possible to use the Killing vector fields in order to derive the angular momentum operators of the underlying geometry \cite{Dowker1974}. 
The application of angular momentum operator enables the access to the preserved angular momentum of the geometry. 
However, here is the very crucial point in the application of these: One can say that the angular momentum operators describe preserved angular momentum if, and only if, the operators describe the global symmetry group of the spacetime. 
The Killing vector fields of the Taub-NUT spacetime in \cref{TN:Eq:KillingVectors} do not describe a global symmetry group, unless the Misner interpretation is considered, i.e. the periodic coordinate time in addition to the coordinate patches of both regular hemispheres. 
Thus, the application of the operators do not contribute to the discussion in the Bonnor interpretation. 
On top of that, this raises the question if a redefinition \cref{TMESol:Ang:Eq:lRule} is reasonable for the Bonnor interpretation at all, since there is no real notion of a orbital angular momentum number. 
Here, we suggest to keep $n$ in the Bonnor interpretation and to use \cref{TMESol:Ang:Eq:lRule} for the Misner interpretation. 
The set to which $l$ belongs will be deduced in the next section.

\subsection{Eigenvalue by the Killing vector fields}
\label{Eigenval:AngMomOp}
The Killing vectors are extended by \enquote{ladder} Killing vector fields
\begin{align}
    \xi_\pm &= \xi_x \pm i \xi_y \notag \\
    &= i e^{\pm i \phi} \left(\pm \frac{\partial}{\partial\theta} + i \cot\theta \frac{\partial}{\partial\phi} + i 2 N (C \cot\theta + \csc\theta) \frac{\partial}{\partial t}\right) \, ,
\end{align}
since any sum of Killing vector fields is again a Killing vector field. 
We note that in order to achieve angular momentum operators, we have to multiply by an additional $-i$, thus:
\begin{subequations}
    \label{Eigenval:AngMomOp:Eq:AngMomOp}
    \begin{align}
        L_t &= -i \xi_t \, , \\
        L_x &= -i \xi_x \, , \\
        L_y &= -i \xi_y \, , \\
        L_z &= -i \xi_z \, , \\
        L_+ &= -i \xi_+ \, , \\
        L_- &= -i \xi_- \, .
    \end{align}
\end{subequations}
For $N = 0$, \cref{Eigenval:AngMomOp:Eq:AngMomOp} become the angular momentum operators for spherically symmetric geometries.
The angular momentum operators consist of derivatives for $t$, $\theta$ and $\phi$.
Therefore, a general function
\begin{align}
    f(t, \theta, \phi) = e^{-i \omega t} e^{i \bar m \phi} S(\theta)
\end{align}
is used for convenience.

From this, the periodicity condition of the $\phi$-coordinate concludes that 
\begin{align}
    m \in \mathbb{Z} \, .
\end{align}
From \cref{Eigenval:AngMomOp:Eq:AngMomOp} squared total angular momentum operators and angular momentum operators of the $z$-axis can be constructed, forming the eigenvalue equations
\begin{subequations}
    \begin{align}
        L^2 f(t, \theta, \phi) &= \tilde \lambda f(t, \theta, \phi) \, , \\
        L_z f(t, \theta, \phi) &= (m - 2 N \omega C) f(t, \theta, \phi) \, ,
    \end{align}
\end{subequations}
where $L^2 = L_x^2 + L_y^2 + L_z^2$.
The eigenvalue of $L_z$ has an additional expression, unlike the $N \rightarrow 0$ case, which is the shift of the azimuthal index $m$ by $2 N \omega C$.
From these properties we can derive a relation of the eigenvalue of $L_z$ and $L^2$, namely
\begin{align}
    (L_x^2 + L_y^2) f(t, \theta, \phi) &= (L^2 - L_z) f(t, \theta, \phi) \, , \label{Eigenval:KV:Eq:lambdarelation}  \\ 
    &= (\tilde \lambda - (m - 2 N \omega C)^2) f(t, \theta, \phi) \, . \notag
\end{align}
The nature of the squared operators dictates that their eigenvalues have to be positive definite, which is why the eigenvalue for $L^2$ has the property of $\tilde \lambda \geqslant 0$. 
This can also be applied to the case of any sum of squared operators.
Together with \cref{Eigenval:KV:Eq:lambdarelation}, this brings up the inequality
\begin{align}
    \tilde \lambda \geqslant (m - 2 N \omega C)^2 \, .\label{Eigenval:AngMomOp:m:Eq:Inequality}
\end{align}
In principle, this leads to the realization that the eigenvalue of $L_z$ is limited by the eigenvalue of $L^2$.
For further convenience we rewrite our equation by $x = \cos\theta$, as in the case of the angular Teukolsky equation. 
Actually evaluating $(L^2 - \tilde \lambda) f(t, \theta, \phi) = 0$, yields
\begin{widetext}
    \begin{align}
        \frac{\partial}{\partial x} \left(\left(1 - x^2\right) \frac{\partial}{\partial x} S(x)\right) - \left(\frac{(m + 2 N \omega (C - 1))^2}{2 (1 + x)} + \frac{(m + 2 N \omega (C + 1))^2}{2 (1 -x)} - \tilde \lambda\right) S(x) &= 0 \, ,     \label{Eigenval:AngMomOp:Eq:DEQ}
    \end{align}
\end{widetext}
which is identical to our angular Teukolsky equation \cref{TMESol:Ang:Eq:DEQ}. 
The actual form of the Eigenvalue is already derived in \cref{Eigenval:Recurrence}.
This identifies 
\begin{align}
    \tilde \lambda = l(l + 1) \, , \label{Eigenval:AngMomOp:Eq:FinalEigenval}
\end{align}
However, by continuing the derivation a restricting property of the free parameters of the spacetime can be found. 
The introduction of the ladder operators becomes relevant now.

Since the commutation of angular momentum operators $L_x$, $L_y$, $L_z$ are all non-vanishing, the same applies to the commutation of
\begin{align}
    [L_z, L_\pm] = \pm L_\pm \, .
\end{align}
This relation can be used to examine the behavior of the ladder operators on eigenfunctions and how the eigenvalue of $L_z$ changes after application by one of the ladder operators. 
We receive: 
\begin{align}
    L_z (L_\pm f(t, \theta, \phi)) = (m - 2 N \omega C \pm 1) (L_\pm f(t, \theta, \phi))
\end{align}
For Taub-NUT there is a contributing term in the ladder operators, unlike, for example, the spherical harmonics. 
Application of $L_\pm$ to the eigenfunctions still change the $L_z$ eigenvalue by $\pm 1$.
As in the usual harmonics, we demand that the application of the ladder operators have to break down at some point.
Expressed mathematically, raising at $m_\text{max}$ and lowering at $m_\text{min}$
\begin{subequations}
    \begin{align}
        L_+ f_{m_\text{max}}(t, x, \phi) &= 0 \, , \\
        L_- f_{m_\text{min}}(t, x, \phi) &= 0 \, ,
    \end{align}
\end{subequations}
should result in vanishing functions.
It is convenient to apply the reversed ladder operators on top of that
\begin{subequations}
    \begin{align}
        L_- L_+ f_{m_\text{max}}(t, x, \phi) &= 0 \, , \\
        L_+ L_- f_{m_\text{min}}(t, x, \phi) &= 0 \, .
    \end{align}
\end{subequations}
A full calculation rewrites $L_\pm L_\mp$ in terms of $L^2$ and $L_z$, of which we know the eigenvalues:
\begin{align}
    L_\pm L_\mp f(t, \theta, \phi) = &\left(L^2 - L_z^2 - L_z\right) f(t, \theta, \phi) \\
    = &\big[\tilde \lambda - (m - 2N \omega C \pm 1)(m - 2 N \omega C)\big] \notag \\
    &\times f(t, \theta, \phi) \notag
\end{align}
Because this expression has to be zero, we have for both cases two different eigenvalues
\begin{subequations}
    \label{Eigenval:AngMomOp:m:Eq:PreEigenval}
    \begin{align}
        \tilde \lambda &= (m_\text{min} - 2 N \omega C) (m_\text{min} - 2 N \omega C - 1) \, , \\
        \tilde \lambda &= (m_\text{max} - 2 N \omega C) (m_\text{max} - 2 N \omega C + 1) \, .
    \end{align}
\end{subequations}
These have to be equal to each other, which yields
\begin{align}
    \Rightarrow &((m_\text{max} - 2 N \omega C) + (m_\text{min} - 2 N \omega C)) \notag \\ 
    &\times (m_\text{min} - m_\text{max} - 1) = 0 \, . \notag
\end{align}
$m_\text{min} - m_\text{max}$ cannot become one, since by definition $m_\text{max} \geqslant m_\text{min}$.
Only the first parenthesis can become zero, leading to
\begin{align}
    \label{Eigenval:AngMomOp:m:Eq:Condition}
    \underbrace{m_\text{max} - 2 N \omega C}_{:= \tilde m_\text{max}} + \underbrace{m_\text{min} - 2 N \omega C}_{:= \tilde m_\text{min}} = 0 \, ,
\end{align}
where we introduced convenient redefinitions $\tilde m_\text{max}$, $\tilde m_\text{min}$.
This can be interpreted that by $C \neq 0$ the set of possible $m$ shifts by $2 N \omega C$. 
It comes in convenient to introduce a transformation $m \mapsto m + 2 N \omega C$.
A first crucial interpretation of the choice of parameters can already be done here: 
\cref{Eigenval:AngMomOp:m:Eq:Condition} requires $2 N \omega C \in \mathbb{Z}$.
This limits the choice of both Taub-NUT relevant parameters together with the frequency.

Using all the information gained until now, \cref{Eigenval:AngMomOp:m:Eq:PreEigenval} turns into
\begin{subequations}
    \label{Eigenval:AngMomOp:m:Eq:Eigenval}
    \begin{align}
        \tilde \lambda &= \tilde m_\text{min} (\tilde m_\text{min} - 1) \\
        \tilde \lambda &= \tilde m_\text{max} (\tilde m_\text{max} + 1)
    \end{align}
\end{subequations}
At this point it is useful to compare the results with the results from \cref{Eigenval:Recurrence}. 
Since the lowest value $l$ can take is $2 N \omega$ (check \cref{TMESol:Ang:Eq:lRule} for $m = 0$ and $n = 0$), \cref{Eigenval:AngMomOp:m:Eq:Eigenval} has to be equal to $l(l+1)$, the Misner condition $2 N \omega \in \mathbb{N}_0$ is derived solely by usage of the angular momentum operators, without introducing any periodicity condition of the Misner interpretation directly.

Applying raising operators $L_+$ to $f_m(t, \theta, \phi)$ from $m_\text{min}$ up to $m_\text{max}$ should result to an integer amount of raising (or lowering) the state.
This can be expressed as as the difference
\begin{align}
    \tilde m_\text{max} - \tilde m_\text{min} &= k \, ,
\end{align}
which should result in $k \in \mathbb{N}_0$. 
Since we know the relation \cref{Eigenval:AngMomOp:m:Eq:Condition}, we receive
\begin{align}
    \Rightarrow m_\text{max} = \frac{k}{2} \, ,
\end{align}
from which we have to conclude that $m \in \mathbb{Z} / 2$.
Comparing \cref{Eigenval:AngMomOp:Eq:FinalEigenval,Eigenval:AngMomOp:m:Eq:Eigenval} leads to the insight that $l \in \mathbb{Z} / 2$ since $\tilde m_\text{max}, \tilde m_\text{min} \in \mathbb{Z} / 2$. 
Only for these values a valid calculation of the linear perturbations by the angular Teukolsky in the Misner interpretation is reasonable.
With this knowledge, the choice of $C$ limits as well, since $2 N \omega C \in \mathbb{Z}$ follows from \cref{Eigenval:AngMomOp:m:Eq:Condition}.
In this case, the Manko-Ruiz parameter only allows for $C \in \{-1, 0, 1\}$.
This result agrees with the insights made a the transformations of \cref{TN:Eq:LocalIsometry} in order to achieve the Misner interpretation of Taub-NUT spacetime. 

Since the angular momentum number $l$ becomes a more reasonable choice for the solutions in the Misner interpretation, this also affects the representation of its completeness relation.
Therefore, the \cref{TMESol:Ang:Eq:CompletenessRelationBonnor} takes the form
\begin{align}
    \sum\limits^\infty_{l = |N\omega|} \sum\limits^l_{\tilde m = -l} \tilde S_{nm}(x) \tilde S_{nm}(x') = \delta(x - x') \label{Eigenval:AngMomOp:Eq:CompletenessRelMisner}
\end{align}

\section{Solutions}
\label{Sol}
\begin{figure*}
    \centering
    \subfloat[$m$ variation, $n = 2$, $N = 1M$, $M\omega = 13$, $C = 0$]{\includegraphics[width=0.4\textwidth]{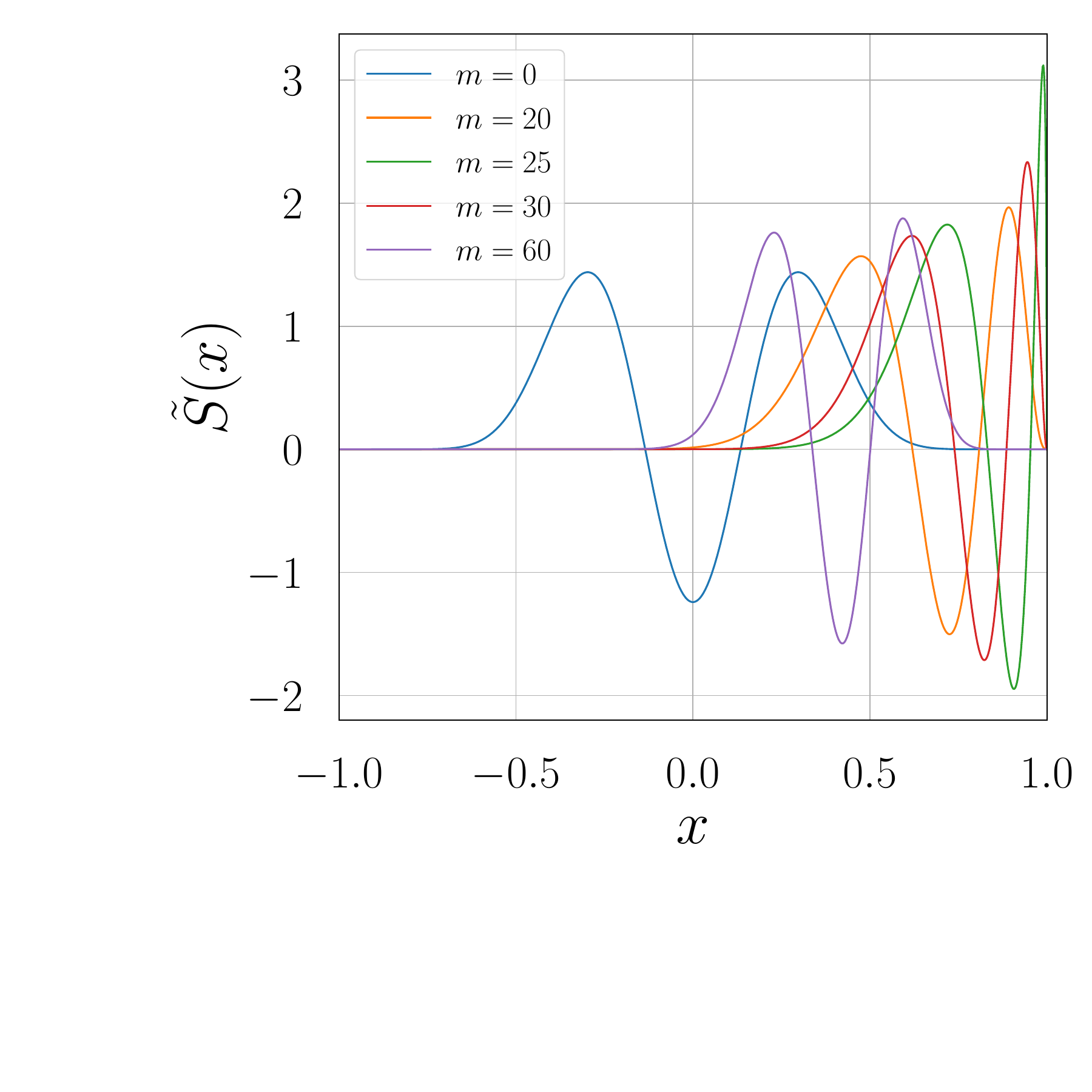}\label{Sol:Fig:SolutionExamples:mvar}} \quad \quad
    \subfloat[$n$ variation, $m = 5$, $N = 1M$, $M\omega = 13$, $C = 0$]{\includegraphics[width=0.4\textwidth]{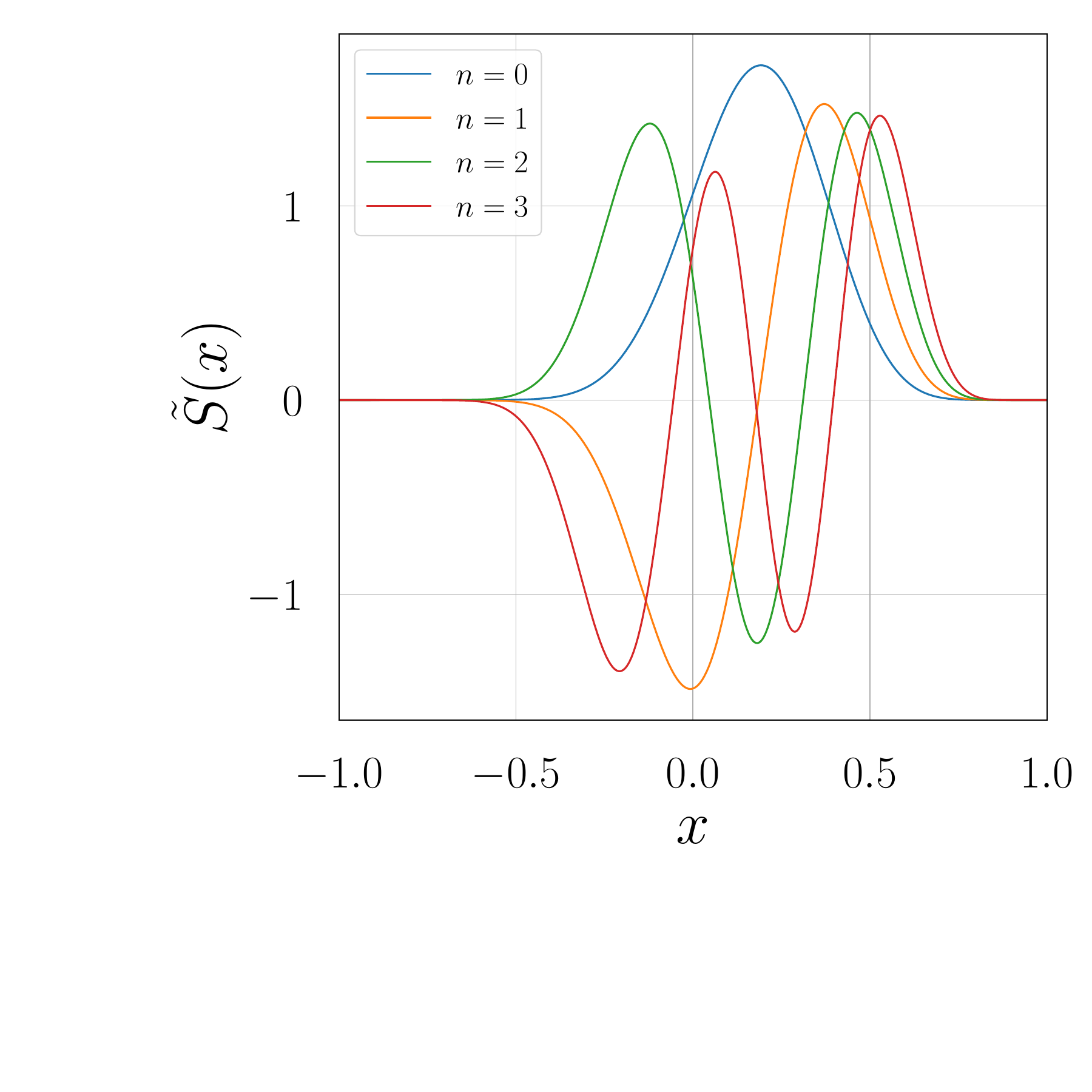}\label{Sol:Fig:SolutionExamples:nvar1}} \\
    \subfloat[$N$ variation, $n = 3$, $m = 0$, $M\omega = 13$, $C = 0$]{\includegraphics[width=0.4\textwidth]{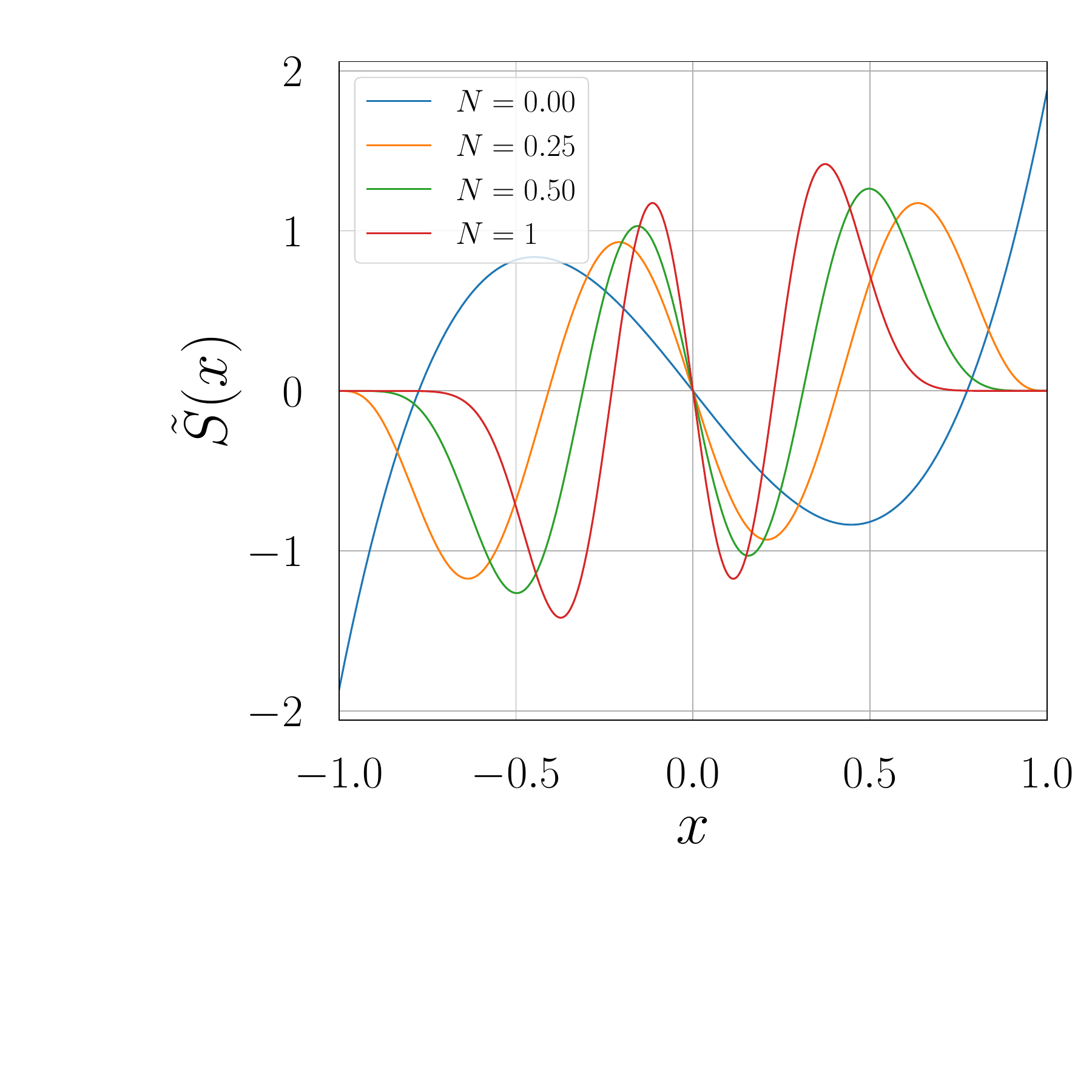}\label{Sol:Fig:SolutionExamples:Nvar1}} \quad \quad
    \subfloat[$N$ variation, $n = 2$, $m = 4$, $M\omega = 13$, $C = 0$]{\includegraphics[width=0.4\textwidth]{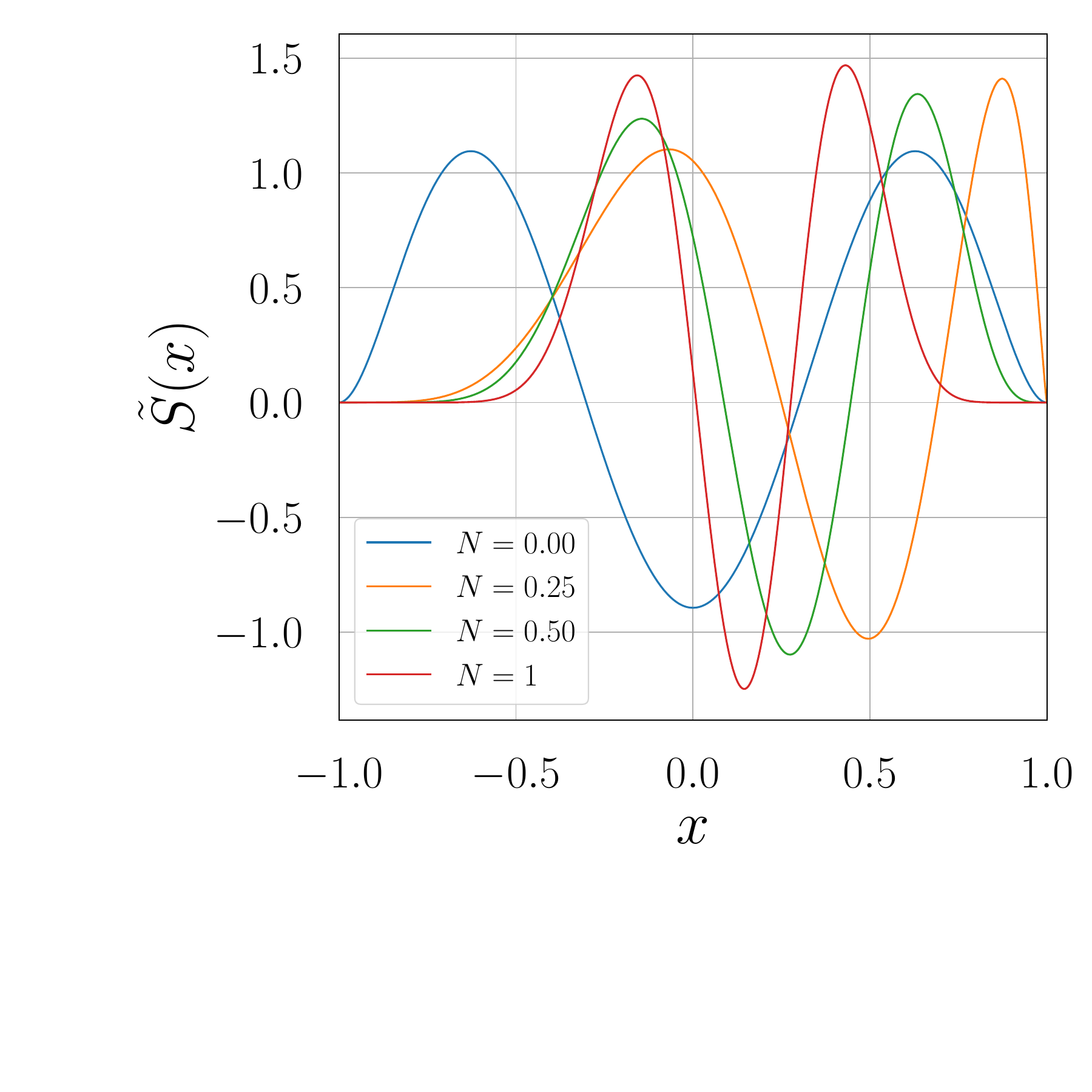}\label{Sol:Fig:SolutionExamples:Nvar2}} \\
    \subfloat[$C$ variation, $n = 3$, $m = 0$, $N = 1$, $M\omega = 13$]{\includegraphics[width=0.4\textwidth]{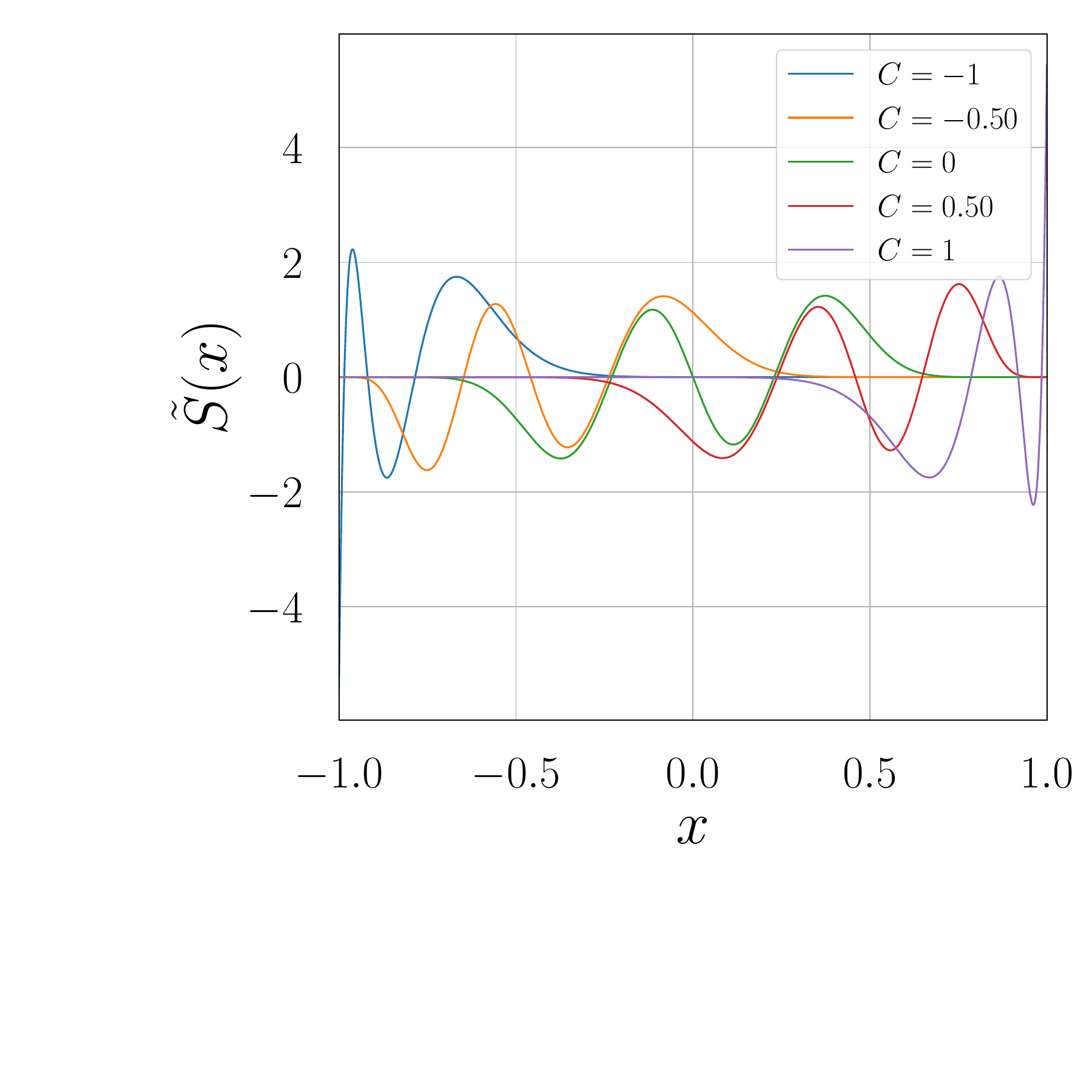}\label{Sol:Fig:SolutionExamples:Cvar}} \quad \quad
    \subfloat[$\omega$ variation, $n = 3$, $m = 0$, $N = 1M$, $C = 0$]{\includegraphics[width=0.4\textwidth]{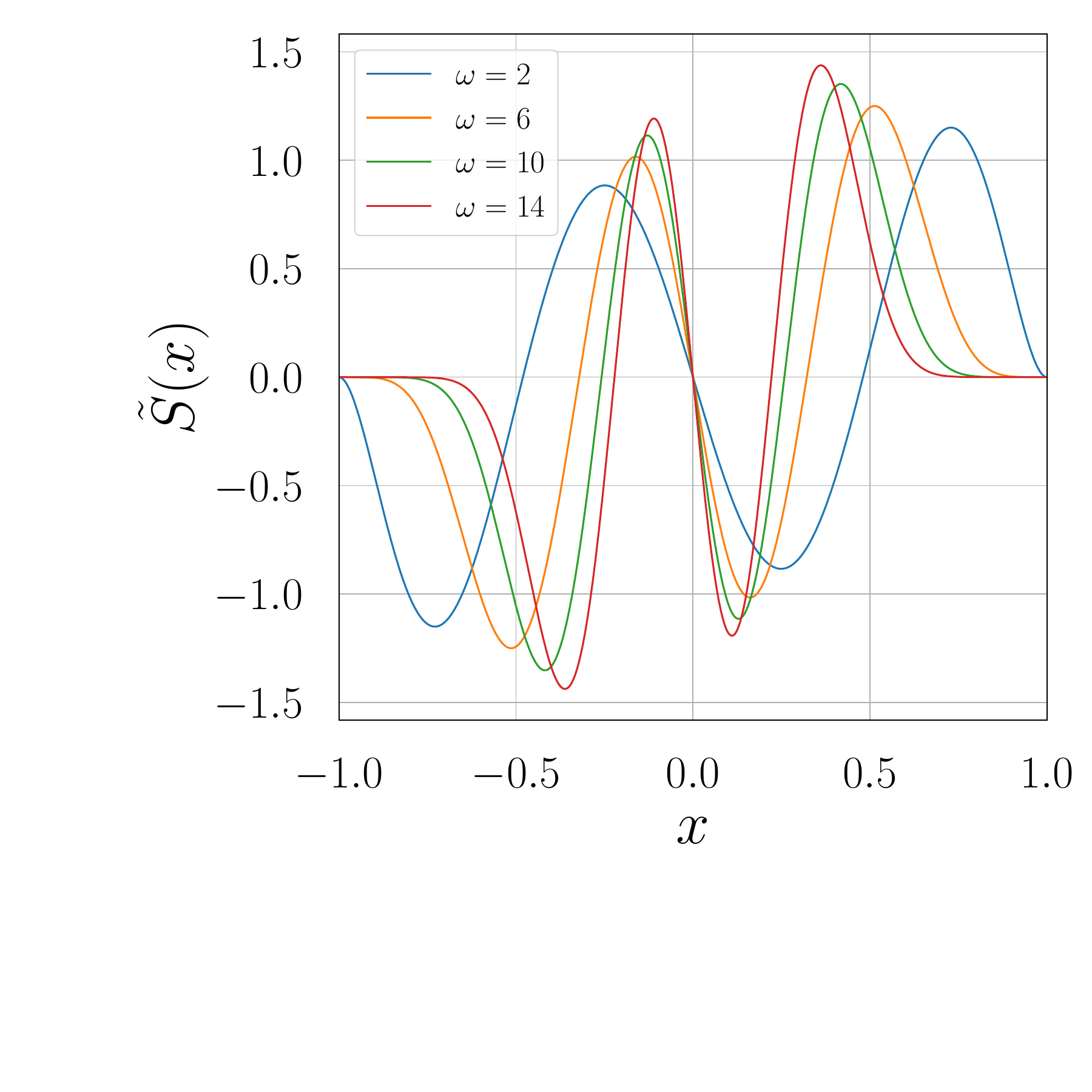}\label{Sol:Fig:SolutionExamples:wvar}}
    \caption{Various evaluations of the solutions of the angular Teukolsky equation. Different parameters are varied depending on the caption of the respective subfigure.}
    \label{Sol:Fig:SolutionExamples}
\end{figure*}

We want to conclude the derivation of the solution of the angular Teukolsky equation for Taub-NUT with some examples and how parameter deviations influence the outcome of them. 
Plots of the solutions can be found in \cref{Sol:Fig:SolutionExamples}, where subfigures will be referenced explicitly in the following. 
The caption of each subfigure lists the respective set of parameters used.
For the moment we keep the $n$ parameter, describing the amount of nodes the harmonics can express and vary $N$ and $\omega$ as real numbers, adopting the idea of the Bonnor interpretation.

A first approach of variation can be examined for the parameter $n$. 
This is shown in \cref{Sol:Fig:SolutionExamples:nvar1}. 
The most prominent consequence is that for $n = 0$ the solution does not express a constant, independent of $x$, as in the spherical harmonics limit $N \rightarrow 0$.
At the poles, this solution becomes zero instead of one. 
Increasing $n$ leads to more nodes for the solution in the same amount of region. 
For $m \neq 0$ and $N \neq 0$, the solution has denser lying node points at a particular range of $x$, losing its previous even distribution.
Compare this to \cref{Sol:Fig:SolutionExamples:Nvar1}. 
In the case of \cref{Sol:Fig:SolutionExamples:nvar1}, these shift to $x = 1$.

\begin{figure}
    \centering
    \includegraphics[width=\linewidth]{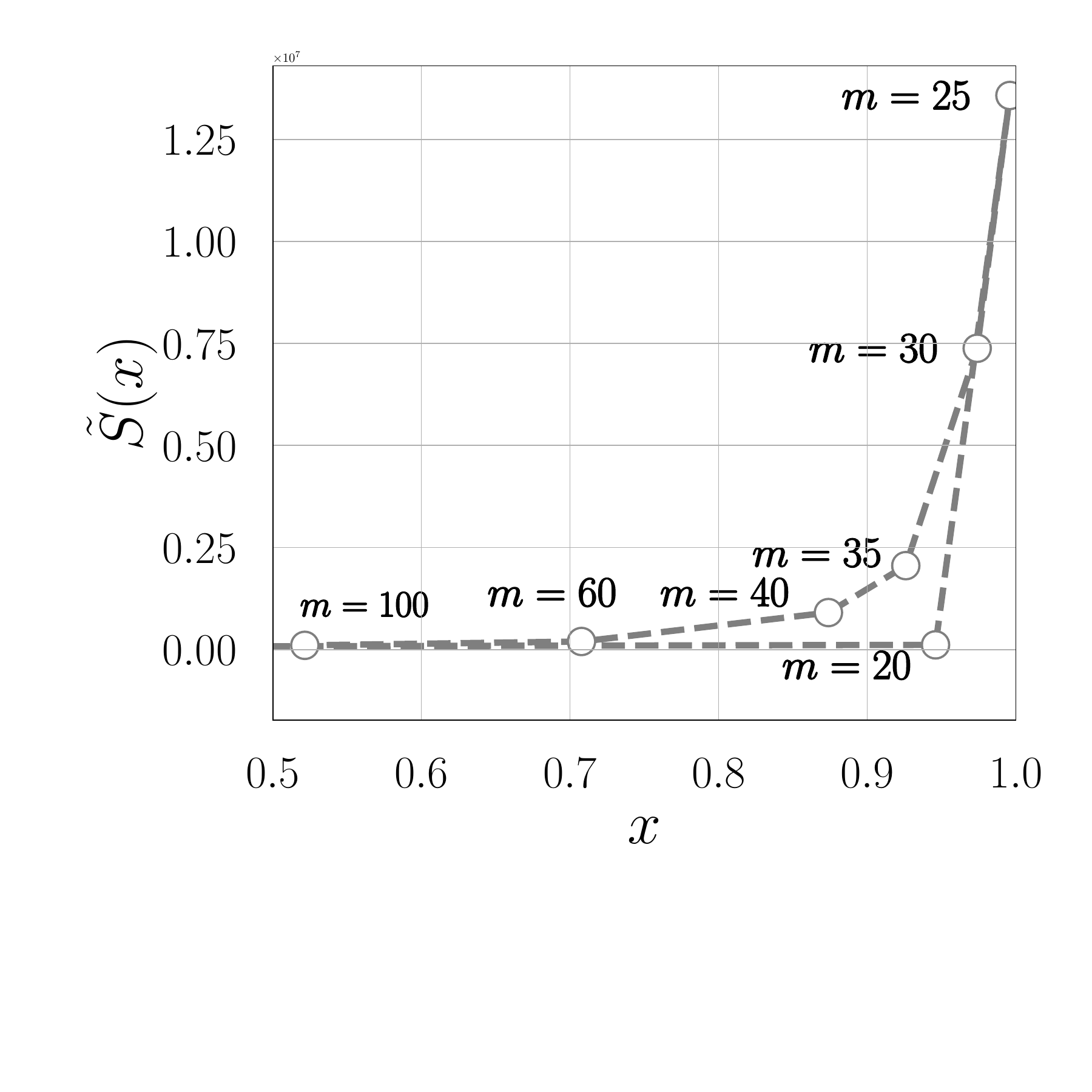}
    \caption{Maximum values of the harmonics for varying $m$. The remaining parameters are equal to the parameter set of \cref{Sol:Fig:SolutionExamples:mvar} except that $n = 2$. Each point is one solution for which the respective $m$ is plotted next to it. The maximum values increase for increasing $m$ up to a certain $m$, after which they decline continuously.}
    \label{Sol:Fig:mvarmaxcurve}
\end{figure}

The variation of $m$ shows for a narrow band of values a shift to $x = 1$, whose magnitude can increase to great values. 
Afterwards, these decline again in magnitude and become smaller. 
This is plotted in a different manner in \cref{Sol:Fig:mvarmaxcurve} for the maximum values of each $m$ and a fixed remaining parameter set. 
This critical $m$, as well as the magnitude of $\tilde S(x)$, shifts to higher values as $N \omega$ increases.

Rather unsurprising are variations in $N$ and $C$. 
Varying $N$ results for $m = 0$ into a squeezing of the oscillations into a narrower region, whereas for $m = 5$ the squeezing is shifted additionally to the south-pole $x = 1$.
For variation in $C$, one can observe a symmetry $S(-x) \mapsto -S(x)$ as $C \mapsto -C$. 
This is expected, since the Misner string switch their magnitudes of induced angular momentum into the spacetime depending on $C$. 
For $|C| = 1$, extreme localization can be observed on the respective poles, whereas for $C = 0$ (and results close to this value) the magnitude of the solutions distribute \enquote{evenly} over $x$.

Finally, the effect of different $\omega$ is similar to the consequences of different $N$, since $N$ and $\omega$ always appear in a product.
Thus, this behavior is to be expected and nothing more of interest can be gained here.

\section{Conclusion and Summary}
\label{Conclusion}
In this work we have discussed the angular Teukolsky equation for the Taub-NUT spacetime. 
The solution of this equation as a second-order ordinary differential equation is given in terms of the confluent Heun function, by which we could show that it in principle reduces to a Confluent Heun polynomial.

The curious properties of the Taub-NUT spacetime enable two different physical interpretations of it. 
The root for different interpretations is the presence of conical singularities outside of the event horizon. 
In one interpretation Misner got rid of the singularities by a two different coordinate patches on both hemispheres and a periodic time coordinate at the expense of closed timelike curves in every event of the spacetime.
The other interpretation of Bonnor accepts the singularities as physical realities analog to Dirac strings of magnetic monopoles. 

The solution of the angular Teukolsky equation in the Bonnor and the Misner interpretations, as well as the corresponding eigenvalues, could be derived exactly analytically.
However, similar to the derivation of (spin-weighted) spherical or spheroidal harmonics, only in the Misner interpretation could the discussion of solutions be extended by the use of the Killing vector fields. 
The reason for this is that the Killing vector fields, acting as generators of the symmetries in our geometry, are only global generators in the Misner interpretation. 
By using the insights of the derivation by means of the recurrence relations and extend them with the knowledge gained in the discussion of the Killing vector fields, one achieved discretised properties of the multipole indices $l$, $m$. 
But most importantly in the case of the Misner interpretation, the Misner condition, i.e. the discretisation of the product $N \omega \in \mathbb{Z}$, was rederived as a consequence of using both approaches. 
Besides, the separation Ansatz in this case can be extended by an additional factor in the exponential function of the $\phi$-component - analog to the case of magnetic monopoles - compensating the additional angular momentum and reducing the differential equations to the same form regardless of the choice of $C$. 
This can be compared with the redefinition of the angular momentum of geodesics in the Taub-NUT spacetime but also appears in a similar manner as in the case of magnetic monopoles \cite{Fierz1944,Tamm1931,Wu1976,Nesterov2008}. 
Let us emphasize again that here a pitfall can be found when using the Killing vector fields: Even though the angular momentum operators with the resulting $L^2$-operator lead to the same angular Teukolsky equation, the insights gained here cannot be used in the Bonnor interpretation as well. 
The Killing vector fields only represent a global generating group in case of the Misner interpretation \cite{Dowker1974,Misner1963,Yohannes2021}. 
However, in the Bonnor interpretation these are only local generators. 
In that case, the preservation of angular momentum is also only locally valid, allowing the examination in the ray-optical approaches, e.g. \cite{Frost2022}.
Interestingly, as in the case of magnetic monopoles, the lower limit of $l$ increases as the product $|N \omega|$ is being increased. 
This raises the question on how the high-frequency limit is being calculated, which should limit against results derived by geodesics (e.g. lensing maps). 

In the Bonnor interpretation, we cannot use these insights and are stuck with the derivation by the recurrence relations. 
And by this, no limitations of the physical parameters can be found. 
The choice of parameters remain free. 

For now we haven't covered the arbitrary spin-weighted cases ($s \neq 0$) due to the fact that for a first approach these weren't necessary. 
We leave the discussion of this for a future work. 
Also the radial part of the Teukolsky equation was not discussed at all, since it does not contribute to the discussed problem at all. 
In a future work, we want to include the radial part as well but also extend to a more general spacetime including a (positive) cosmological constant (i.e. Taub-NUT-de Sitter). 
We want to contribute the results obtained here to a future work where the scattering can be solved analytically by means of the general Heun equation and wave-optical effects at an observer (cf. \cite{Motohashi2021,Willenborg2024,Kubota2024,Hatsuda2020}).

\section{Acknowledgments}
\label{Ack}
The authors would like to thank Volker Perlick, Christian Pfeifer, Domenico Giulini and Torben Frost for insightful discussions.
Additionally, gratitude is expressed for the authors of the software package {\texttt{xAct}} \cite{xAct} for Mathematica, which was used in the calculation of the Teukolsky master equation 
The research is funded by the Deutsche Forschungsgemeinschaft (DFG, German Research Foundation) – Project-ID 434617780 – SFB 1464 and
funded by the Deutsche Forschungsgemeinschaft (DFG, German Research Foundation) under Germany’s Excellence Strategy – EXC-2123 QuantumFrontiers – 390837967, and through the Research Training Group 1620 \enquote{Models of Gravity}.

\section{Appendix}
\label{App}
\appendix
\section{Plebański-Demiański}
\label{App:PB}
The Plebański–Demiański metric, as the most general axial-symmetric and stationary metric for electrovacuum solutions of compact bodies in general relativity of Petrov Type D, is defined as \cite{Griffiths2012,Grenzebach2016}
\begin{align}
    ds^2 =~ &\frac{1}{\Omega^2} \bigg(\frac{1}{\Sigma} \left(\Delta_r - a^2 \Delta_\theta \sin^2\theta \right) dt^2 \label{TN:Eq:PBmetric} \\
    &+ \frac{2}{\Sigma} \left(\Delta_r \chi - a (\Sigma + a \chi) \Delta_\theta \sin^2 \theta\right) dt d\phi \notag \\
    & - \frac{1}{\Sigma} \left[(\Sigma + a \chi)^2 \Delta_\theta \sin^2 \theta - \Delta_r \chi^2 \right] d\phi^2 \notag \\
    & - \frac{\Sigma}{\Delta_r} dr^2 - \frac{\Sigma}{\Delta_\theta} d\theta^2\bigg) \notag \, ,
\end{align}
where its metric functions are
\begin{subequations}
    \begin{alignat}{2}
        \Omega &=: \Omega(r, \theta) &&= 1 - \frac{\alpha_P}{\omega_P} (N + a \cos\theta) r \, , \\
        \Sigma &=: \Sigma(r, \theta) &&= r^2 + (N + a \cos\theta)^2 \, , \\
        \chi &=: \chi(\theta) &&= a \sin^2\theta - 2 N (\cos\theta + C) \, , \\
        \Delta_\theta &=: \Delta_\theta(\theta) &&= 1 - a_3 \cos\theta - a_4 \cos^2\theta \, , \\
        \Delta_r &=: \Delta_r(r) &&= b_0 + b_1 r + b_2 r^2 + b_3 r^3 + b_4 r^4 \, .
    \end{alignat}
\end{subequations}
The coefficients of the $\Delta_\theta$ and $\Delta_r$ polynomials are  
\begin{subequations}
    \begin{align}
        \label{TN:Eq:DeltaThetaPBCoeff}
        a_3 =& a \left(2 \frac{\alpha_P}{\omega_P} M - 4 N \left[\frac{\alpha_P^2}{\omega_P^2} (k + \beta) + \frac{\Lambda}{3}\right]\right) \, , \\
        a_4 =& -a^2 \left(\frac{\alpha_P^2}{\omega_P^2} (k + \beta) + \frac{\Lambda}{3}\right) \, ,
    \end{align}
\end{subequations}
and
\begin{widetext}
    \begin{subequations}
        \label{TN:Eq:DeltarPBCoeff}
        \begin{align}
            b_0 =& k + \beta \, , \\
            b_1 =& -2 M \, , \\
            b_2 =& \frac{k}{a^2 - N^2} + 4 \frac{\alpha_P}{\omega_P} N M - \left(a^2 + 3 N^2\right) \left(\frac{\alpha_P^2}{\omega_P^2} (k + \beta) + \frac{\Lambda}{3}\right) \, , \\
            b_3 =& -2 \frac{\alpha_P}{\omega_P} \left[ \frac{k N}{a^2 - N^2} - \left(a^2 - N^2\right) \left(\frac{\alpha_P}{\omega_P} M - N\left(\frac{\alpha_P^2}{\omega_P^2} (k + \beta) + \frac{\Lambda}{3}\right)\right) \right] \, , \\
            b_4 =& -\left(\frac{\alpha_P^2}{\omega_P^2} k + \frac{\Lambda}{3}\right) \, ,
        \end{align}
    \end{subequations}
\end{widetext}
including the abbreviated definitions $k$, $\omega_P$, and $\beta$, appearing in \cref{TN:Eq:DeltaThetaPBCoeff,TN:Eq:DeltarPBCoeff}
\begin{subequations}
\begin{align}
k &= \frac{1 + 2 \frac{\alpha_P}{\omega_P} N M - 3 N^2 \left(\frac{\alpha_P^2}{\omega_P^2} \beta + \frac{\Lambda}{3}\right)}{1 + 3 \frac{\alpha_P^2}{\omega_P^2} N^2 \left(a^2 - N^2\right)} \, , \\
\omega_P &= \sqrt{a^2 + N^2} \, .
\end{align}
\end{subequations}
Besides the aforementioned parameters of interest, the Plebański–Demiański metric also contains the charge parameter $\beta$ comprising the electric and magnetic charge, the cosmological constant $\Lambda$, the Kerr spin parameter $a$ and the acceleration parameter $\alpha_P$.

\section{On the TME for Taub-NUT}
\label{App:NP}
The discussion of linear perturbations of a background metric is intuitively done by considering the decomposition $g_{\mu\nu} = \eta_{\mu\nu} + h_{\mu\nu}$.
It is used to study, e.g., the quasinormal modes of black holes \cite{Zerilli1970}. 
Another approach to the perturbation problem was taken by Teukolsky \cite{Teukolsky_1973} via the Newman-Penrose (NP) formalism. 
The advantage of perturbing the metric in the NP formalism is the extension to fields with an arbitrary spin.
The 12 NP scalars form the basis of the formalism, and their derivation is performed in terms of a null tetrad system. 
The choice of the two real null tetrads $l$ and $n$, representing radial in- and outgoing null rays viewed from an asymptotic region, and a complementary complex null tetrad $m$ with its complex conjugate $\bar m$ is substantial. 
They fulfill
\begin{subequations}
    \begin{align}
        l_a n^a &= -1, \\
        m_a \bar{m}^a &= 1,
    \label{TME:KT:Eq:NonZero}
    \end{align}
\end{subequations}
which are the only non-zero contractions.
In Petrov type-D metrics, some degrees of freedom vanish, leading to $\kappa = \sigma = \lambda = \nu = \pi = \tau = 0$. 
However, a complex null rotation of one of the null tetrads
\begin{align}
    m \mapsto e^{i \psi(r, \theta)} m
\end{align}
is not yet fixed in order to get a 6-parameter group of homogeneous Lorentz transformation which preserve the tetrad orthogonality relation \cite{Teukolsky_1973}. 
Demanding that $\epsilon \overset{!}{=} 0$ also fixes the last degree of freedom in this group, leading to the Kinnersley tetrads \cite{Kinnersley_1969}, by which
\begin{align}
    \psi(r, \theta) = \arctan \frac{r}{N} + c_0 \, ,
\end{align}
where $c_0 = 0$.
The purpose of this step is to achieve well behaved tetrads on the event horizon. 
For the Taub-NUT metric, expressed in PB functions, this yields
\begin{subequations}
\begin{align}
    l^\mu &= \frac{1}{\Delta_r} \left(\Sigma \partial_t + \Delta_r \partial_r \right], \\
    n^\mu &= \frac{1}{2 \Sigma} \left[\Sigma \partial_t - \Delta_r \partial_r \right], \\
    m^\mu &= -\frac{i \rho^*}{\sqrt{2}} \left[\chi \csc\theta \partial_t - i \partial_\theta + \csc\theta \partial_\phi \right].
\end{align}
\end{subequations}
The remaining non-zero coefficients are
\begin{subequations}
\begin{align}
    \rho &= -\frac{1}{r - i N} \label{App:NP:Eq:rho} \\
    \mu &= \frac{\rho \Delta_r}{2 \Sigma} \\
    \gamma &= \frac{2 \Delta_r \rho + \Delta_r'}{4 \Sigma} \\
    \beta &= -\frac{\rho^* \cot\theta}{2 \sqrt{2}} \\
    \alpha &= \frac{\rho \cot\theta}{2 \sqrt{2}}
\end{align}
\end{subequations}
From these, 5 complex Weyl scalars $\Psi_i$ can be composed. 
Type-D spacetimes eliminate certain scalars ($\Psi_1 = \Psi_3 = \Psi_4 = \Psi_5 = 0$).  
The remaining non-zero Weyl scalar is
\begin{align}
    \Psi_2 = (M - i N) \rho^3 \, . 
\end{align}
In contrast to linear perturbation in the metric, all spin coefficients, Kinnersley tetrads and Weyl scalars are linearly perturbed. 
From these perturbations and related symmetries, the differential equations for different spin weights are derived in terms of the NP formalism \cite{Bini2003}, where for scalar perturbations ($s = 0$)
\begin{align}
    [&D\Delta + \Delta D - \delta^* \delta - \delta \delta^* + (-\gamma - \gamma^* + \mu + \mu^*) D \notag \\
    & + (\epsilon + \epsilon^* - \rho^* - \rho) \Delta + (-\beta^* - \pi + \alpha + \tau^*) \delta \\
    & + (-\pi^* + \tau - \beta + \alpha^*) \delta^*] {}_s\Psi_{lm} = 0 \, . \notag
\end{align}
The resulting differential equation for the scalar case coincides with the Klein-Gordon equation $\Box \Phi = 0$. 
For positive spin-weights $\left(s \in \left\{\frac{1}{2}, 1, 2\right\}\right)$ the Teukolsky master equation follows,
\begin{align}
    \{&[D - \rho^* + \epsilon^* + \epsilon - 2 s (\rho + \epsilon)] \left(\Delta + \mu - 2 s \gamma\right) \notag \\
    & - [\delta + \pi^* - \alpha^* + \beta - 2 s (\tau + \beta)] (\delta^* + \pi - 2 s \alpha) \\
    & - 2 (s - 1) (s - 1/2) \psi_2\} {}_s\Psi_{lm} = 0 \notag
\end{align}
and for negative spin-weights $\left(s \in \left\{-\frac{1}{2}, -1, -2\right\}\right)$ we obtain
\begin{align}
    \{&[\Delta - \gamma^* + \mu^* - \gamma - 2 s (\gamma + \mu)] (D - \rho - 2 s \epsilon) \notag \\
    & - [\delta^* - \tau^* + \beta^* - \alpha - 2 s (\alpha + \pi)] (\delta - \tau - 2 s \beta) \\
    & - 2 (s + 1)(s + 1/2) \psi_2\} {}_s\Psi_{lm} = 0 \, , \notag
\end{align}
where $D = l_\mu \partial^\mu$, $\Delta = n_\mu \partial^\mu$, and $\delta = m_\mu \partial^\mu$ are directional derivatives of the NP formalism. 
Another noteworthy extension proceeds to supersymmetric spin fields $s = \pm 3/2$, which is approached via the Geroch-Held-Penrose formalism \cite{Geroch1973}. 

The solution ${}_s\Psi_{lm}$ of the differential equation yields scalars of the Newman-Penrose or Geroch-Held-Penrose formalism according to \cref{TME:Tab:NPGHPsol}.
\begin{table}[t!]
    \centering
    \begin{tabular}{c||c|c|c|c|c|c|c|c|c}
        $s$ & 0 & $\frac{1}{2}$ & $-\frac{1}{2}$ & 1 & -1 & $\frac{3}{2}$ & -$\frac{3}{2}$ & 2 & -2 \\
        \hline
        ${}_s\Psi_{lm}$ & $\Phi$ & $\chi_0$ & $\rho^{-1} \chi_1$ & $\phi_0$ & $\rho^{-2} \phi_2$ & $\Omega_0$ & $\rho^{-3} \Omega_3$ & $\psi_0^B$ & $\rho^{-4} \psi_2^B$
    \end{tabular}
    \caption{Solution of ${}_s\Psi_{lm}$ for different spin-weights $s$. The respective expressions are scalars from Newman-Penrose or rather Geroch-Heldt-Penrose formalism \cite{Teukolsky_1973}.}
    \label{TME:Tab:NPGHPsol}
\end{table}
Finally, for the Taub-NUT metric using the NP identity \cite{Chandrasekhar1998}
\begin{align}
    D\mu - \delta\pi = &(\rho^* \mu + \sigma \lambda) + \pi \pi^* - (\epsilon + \epsilon^*) \mu - (\alpha^* - \beta)\pi \notag \\
    & - \nu \kappa + \psi_2 \, .
\end{align}

%

\end{document}